\documentclass{article}
\usepackage{graphicx} 
\usepackage{amsmath}
\usepackage[letterpaper, total={7in, 9.5in}]{geometry}
\title{turbo-RANS: Straightforward and Efficient Bayesian Optimization of Turbulence Model Coefficients}
\author{Ryley McConkey$^{\text{a}*}$, Nikhila Kalia$^\text{a}$, Eugene Yee$^\text{a}$, Fue-Sang Lien$^\text{a}$}
\date{\small $^\text{a}$ University of Waterloo, Canada.\\ 200 University Ave W, Waterloo, ON N2L 3G1\\
$^*$ Corresponding author: rmcconke@uwaterloo.ca
}
\DeclareUnicodeCharacter{2212}{-}
\usepackage{caption}
\usepackage{xcolor}
\usepackage{subcaption}
\usepackage{newtxtext}
\usepackage{newtxmath}
\usepackage{url}
\usepackage[sorting=nyt,dashed=false,style=authoryear-icomp,maxcitenames=2,maxbibnames=99,natbib]{biblatex}
\usepackage{lineno}
\let\svthefootnote\thefootnote
\newcommand\freefootnote[1]{%
  \let\thefootnote\relax%
  \footnotetext{#1}%
  \let\thefootnote\svthefootnote%
}

\DefineBibliographyStrings{english}{
  andothers = {\mkbibemph{et\addabbrvspace al\adddot}}
}
\usepackage{sectsty}
\subsectionfont{\normalfont\itshape}
\subsubsectionfont{\normalfont\itshape}
\usepackage{hyperref}

\begin{document}

\maketitle
\section*{Abstract}
\textbf{Purpose.} Industrial simulations of turbulent flows often rely on Reynolds-averaged Navier-Stokes (RANS) turbulence models, which contain numerous closure coefficients that need to be calibrated. In this work, we address this issue by proposing a semi-automated calibration of these coefficients using a new framework (referred to as turbo-RANS) based on Bayesian optimization. \\
\textbf{Design/methodology/approach.} We introduce the generalized error and default coefficient preference (GEDCP) objective function, which can be used with integral, sparse, or dense reference data for the purpose of calibrating RANS turbulence closure model coefficients. Then, we describe a Bayesian optimization-based algorithm for conducting the calibration of these model coefficients. An in-depth hyperparameter tuning study is conducted to recommend efficient settings for the turbo-RANS optimization procedure. \\
\textbf{Findings.} We demonstrate that the performance of the $k$-$\omega$ shear stress transport (SST) and Generalized $k$-$\omega$ (GEKO) turbulence models can be efficiently improved via turbo-RANS, for three example cases: predicting the lift coefficient of an airfoil; predicting the velocity and turbulent kinetic energy fields for a separated flow; and, predicting the wall pressure coefficient distribution for flow through a converging-diverging channel. \\
\textbf{Originality.} This work is the first to propose and provide an open-source black-box calibration procedure for turbulence model coefficients based on Bayesian optimization. We propose a data-flexible objective function for the calibration target. Our open-source implementation of the turbo-RANS framework includes OpenFOAM, Ansys Fluent, STAR-CCM+, and solver-agnostic templates for user application. \\
\textbf{Keywords. }Turbulence modelling, Bayesian optimization, calibration, data-driven methods, Reynolds-averaged Navier Stokes, GEKO model.\\
\textbf{Paper type.} Research paper.
\section{Introduction}

Turbulent flows are found in many engineering applications such as automotive, aerospace, chemical, nuclear, and wind energy. Engineering activities such as design, optimization, safety assessment, maintenance, and failure investigation in these industries frequently require predicting the behaviour of turbulent flows. Predicting turbulent flow is a necessary, but complex problem, that is often addressed using experimental or computational approaches. While experimental approaches often produce the most physically accurate results, they are expensive and limited in the quantities of interest (information) they can provide. For these reasons, predictions obtained using computational fluid dynamics (CFD) are attractive due to their low cost, capability to rapidly explore a (frequently large) design space, and ability to access detailed information about practically all aspects of the flow.

Many industrial CFD simulations rely on solving the Reynolds-averaged Navier-Stokes (RANS) equations, which are more computationally affordable than conducting large-eddy simulation (LES)~\citep{Witherden2017}. RANS models have been considered an industrial workhorse for the past several decades. However, there are several well-known deficiencies with the application of these methods. A major source of error in RANS is the turbulence model, which often relies on a single-point, linear-eddy-viscosity closure model (e.g., turbulence closure models such as $k$-$\varepsilon$~\citep{Launder1974}, $k$-$\omega$~\citep{Wilcox1988}, Spalart-Allmaras~\citep{Spalart1992}). The past several decades of experience with RANS have resulted in the development of sophisticated turbulence closure models such as the $k$-$\omega$ shear stress transport (SST) model~\citep{Menter1994}, the $k$-$\varepsilon$-$\phi$-$f$ model~\citep{Laurence2005}, and the recently-developed generalized $k$-$\omega$ (GEKO) model~\citep{GEKO_model}. However, structural limitations that arise as a result of the assumptions and approximations underpinning single-point RANS closure models affect their predictive performance for in various engineering and industrial applications~\citep{Wilcox1994,Pope2000}.

Recently, there has been widespread interest in using data-driven approaches to improve and calibrate RANS models~\citep{Brunton2020,Duraisamy2019a}. Because of the ability to automate traditional physics-based and heuristic turbulence modeling, more advanced models can be potentially developed. Machine learning has been used to develop complex non-linear eddy viscosity models, which can provide good approximations for highly-resolved LES and direct numerical simulation (DNS) closure terms for use with the RANS modelling framework for incompressible flows~\citep{Ling2016,McConkey2022,Kaandorp2020,Wu2018}, and compressible flows~\citep{Grabe2023,Sciacovelli2023}. Additionally, machine learning can be used to generate LES subgrid closure models~\citep{Maulik2019} and accelerate DNS through the application of super-resolution models~\citep{Kochkov2021}. While some approaches aim to use highly complex machine learning models to replace or augment the turbulence closure relation, another promising approach is using reference data to calibrate closure coefficients in an existing turbulence model. Most RANS turbulence closure models contain a number of tuneable coefficients, which can be calibrated to improve the RANS predictive performance for certain classes of flows. The problem of calibrating the turbulence model coefficients so that a prediction is in better conformance with the reference data can be formulated as an optimization problem. To this purpose, a number of previous studies have proposed the use of genetic algorithms~\citep{Gimenez2019,Mauro2017}, Nelder-Mead simplex algorithm~\citep{Yoder2021}, optimal Latin hypercube method~\citep{Gu2012}, and the global search method~\citep{Barkalov2022} for addressing this problem. Bayesian optimization was demonstrated to be an effective method for calibrating model coefficients in multiscale dynamical systems by ~\citet{Nadiga2019}. As discussed by ~\citet{Diessner2022}, the Bayesian optimization methodology is particularly promising for addressing many fluid mechanics problems involving model coefficient calibration because it is efficient at optimizing expensive functions. For example, ~\citet{Morita2022} demonstrated the efficient performance of Bayesian optimization for applications to shape optimization and to optimal icing control problems. Additionally, zonal coefficient optimization can lead to improved predictive performance, as demonstrated by the zonal eddy viscosity approach of~\citep{Matai2019}.

Naturally, different optimization problems will require different optimization techniques. Some techniques are more suited for high-dimensional problems, noisy problems, or problems where the gradient of the objective function is analytically available. In our view, the primary consideration for the turbulence model calibration problem is the computational cost associated with evaluating the objective function. If we formulate the objective function as an error measure between our reference data and the associated predictions of this data for a given set of model coefficients, then every evaluation of the objective function will require a new CFD calculation using the new turbulence model coefficients. The cost of a CFD calculation varies with the specific problem, but for an industrially relevant flow, a single three-dimensional (3D) RANS calculation can easily require on the order of hundreds to thousands of central processing unit (CPU) hours. Therefore, this optimization problem demands a technique that can carefully and efficiently select new query points (viz., new values for the tuneable model coefficients at which the evaluation of the predictive performance will occur next). Bayesian optimization is a popular choice for optimizing computationally expensive objective functions. For example, Bayesian optimization is commonly used in neural network hyperparameter tuning, where each evaluation of the objective function typically involves an expensive training run~\citep{Garnett_BayesOpt_Book}.

Bayesian approaches have been applied for calibrating turbulence models in a number of previous studies. In particular, Bayesian inference has received attention because it can provide information on uncertainty in a RANS model prediction. ~\citet{Maruyama2021} applied Bayesian inference to infer model coefficients of the Spalart-Allmaras (SA) model for a transonic flow over a wing. SA model coefficients were also inferred for several transonic airfoil problems in an investigation conducted by ~\citet{DaRonch2020}, who found that a properly calibrated set of model coefficients improved the predictive accuracy on transonic test cases outside of the reference data used for the calibration. ~\citet{Ray2018} developed an analytical model for the optimal coefficients of the $k$-$\varepsilon$ turbulence model for a jet in a cross flow and also performed Bayesian calibration of the turbulence model coefficients. In particular, these investigators hypothesized that for many flows, model-form errors in RANS may be significantly smaller than coefficient calibration errors. A similar investigation, involving the calibration of the SA model for a jet flow with a concomitant determination of the uncertainty in the tuned model coefficients, was undertaken by ~\citet{Li2022}. ~\citet{Edeling2014} analyzed the error in the Launder-Sharma $k$-$\varepsilon$ turbulence model for wall-bounded flows and applied Bayesian inference for calibrating the model coefficients. An earlier investigation conducted by ~\citet{Oliver2011} applied Bayesian uncertainty quantification to four eddy-viscosity models. Recently, ~\citet{Fischer2023} coupled a neural network with Bayesian optimization to predict optimal values of coefficients of the generalized $k$-$\omega$ (GEKO) model~\citep{GEKO_model} for a trench-shaped film-cooling flow. Another popular approach is the field inversion machine learning approach proposed by ~\citet{Duraisamy2017_fieldinversion} which is often applied to infer local values of the calibration coefficients.

In this work, we introduce the {\bf tu}ning pa{\bf r}ameters using {\bf B}ayesian {\bf o}ptimization{\bf -RANS} (turbo-RANS) platform. More specifically, turbo-RANS is a framework for applying Bayesian optimization to the RANS coefficient calibration problem. Bayesian optimization is particularly suited to problems where an evaluation of the objective function is computationally expensive---this is certainly the case for the turbulence model calibration problem in RANS. We formulate an objective function for use with turbo-RANS that is versatile in terms of the types of reference data that may be used. For three different model coefficient calibration problems, we demonstrate that turbo-RANS can be applied to tune turbulence model coefficients in a computationally efficient manner. Another contribution of the present work is the provision of an open-source implementation of our proposed calibration framework. We illustrate the application of this framework on several example codes. Templates and source code have been made available in a GitHub repository~\citep{turborans_github}. We also investigate how the hyperparameters associated with the Bayesian optimization can be chosen and, in the course of this investigation, we provide general recommendations for the settings of these various hyperparameters that lead to a computationally efficient and stable calibration of the RANS model coefficients.

A number of previous studies have applied a Bayesian framework to the turbulence model coefficient calibration problem. However, most of these previous studies have focused on the problem as an inference problem. For example, given a set of RANS calculations and some reference data, Bayesian inference can provide a best estimate plus uncertainty (BEPU) of the turbulence closure coefficients. However, this methodology is computationally prohibitive for use with very expensive RANS computations (forward model predictions). Indeed, the computational demands of the RANS model predictions make it practically impossible as a tool for sampling from the posterior distribution using Markov chain Monte Carlo (MCMC) to produce a sequence of dependent draws from the posterior. In the case where one is under the constraint of a very limited computational budget (typically less than about 100 RANS model calculations), we advocate the application of Bayesian optimization to the problem: given previous knowledge of the problem, what is the next most efficient point to query? The use of the Bayesian optimization approach here results in a naturally computationally efficient optimization procedure.

The Bayesian optimization approach allows the exploration and exploitation tradeoff to be tightly controlled during optimization. Moreover, the objective function proposed here is flexible and pragmatic and, as a result, is well-suited for use with Bayesian optimization. Whereas many previous studies used simple objective functions (e.g., root-mean-squared error), the objective function proposed here produces a well-posed optimization problem and can simultaneously be used with different types of reference data. We demonstrate that using this objective function for calibration leads generally to an accurate prediction of various quantities of interest derived from the turbulent flow. Finally, we believe that the complexity of the implementation and the ease of use (or lack thereof) are often overlooked factors that limit the widespread use of Bayesian calibration. To address these factors, we present a straightforward methodology and provide an open-source implementation that can be immediately used in a variety of CFD contexts.

This paper is organized as follows. In Section~\ref{sec:methodology}, we highlight relevant concepts of Bayesian optimization. We then discuss the formulation of the generalized objective function used in the turbo-RANS framework, providing a rationale for this formulation. The turbo-RANS framework is then described in detail along with how this framework is implemented. In Section~\ref{sec:demonstrations}, we apply turbo-RANS to calibrate coefficients in the $k$-$\omega$ SST and GEKO turbulence closure models for three different types of reference data: namely, integral parameter data, dense DNS data, and sparse wall measurement data. In Section~\ref{sec:hyperparameters}, we perform an extensive study of how to choose the hyperparameters associated with the Bayesian optimization used in the RANS model calibration. Based on this hyperparameter investigation, we provide recommended settings for the hyperparameters that appear in the Bayesian optimization schema, including those used in the Gaussian process regression. Finally, Section~\ref{sec:conclusion} provides a conclusion---here, we contextualize the results obtained in this study and provide suggestions for future work.

\section{Background and Methodology}\label{sec:methodology}
This section is organized as follows. Sections~\ref{sec:bo}--\ref{sec:utility} provide a high-level overview of Bayesian optimization. Section~\ref{sec:gedcp} explains the proposed generalized error and default coefficient preference (GEDCP) objective function, which we recommend for optimizing turbulence model coefficients. Section~\ref{sec:algorithm} provides details and a discussion of the turbo-RANS framework. Implementation of this framework is discussed in Section~\ref{sec:implementation}.

\subsection{Bayesian optimization}\label{sec:bo}
We summarize the main ideas underpinning Bayesian optimization and Gaussian process regression as they are applied in the turbo-RANS framework. Bayesian optimization is described in detail in~\citet{Garnett_BayesOpt_Book}, and Gaussian process regression is explained in detail in~\citet{Rasmussen_GP_Book}.

The goal of Bayesian optimization is to maximize an objective function $f(x)$ with respect to a vector of parameters $x$. Since this objective function can be computationally expensive to compute, we create a surrogate model of the objective function that is significantly less expensive to compute. It is useful for the surrogate model to return both the best estimate $\mu(x)$ of the objective function and the uncertainty $\sigma(x)$ in this best estimate. We introduce a utility function $\alpha(x) \equiv \alpha(\mu(x),\sigma(x))$ that combines the information embodied in the predicted objective function value and uncertainty in this value into a single scalar-valued function (see Section~\ref{sec:utility}). The next $x$ value queried is the point that maximizes the utility function $\alpha(x)$. This utility function allows us to control the tradeoff between exploration and exploitation. Exploration is the evaluation of the objective function in new regions far away from the currently available data (i.e., points with high uncertainty $\sigma(x)$). Exploitation is the evaluation of the objective function in regions where it is expected to be favourable based on past data (i.e., points with a high expected value of $\mu(x)$). The primary advantage of Bayesian optimization is the ability to control this exploration/exploitation tradeoff, which produces an optimization procedure that is efficient in terms of the number of objective function evaluations required to find the optimal value of $x$.
\subsection{Gaussian process regression}\label{sec:gp}
In Bayesian optimization, Gaussian process regression is typically used to construct a surrogate model of the objective function. Given a dataset $\mathcal{D}$ consisting of a set of values of the objective function $f(x)$ at points $x$ ($\bigl\lbrace(x_1,f(x_1)), (x_2, f(x_2)), \ldots, (x_n,f(x_n))\bigr\rbrace$), we can fit a Gaussian process regression model $\mathcal{GP}(x)$ and use it as a surrogate model for $f(x)$. A Gaussian process regression model fits a Gaussian process to $f(x')$ for all $x'$ in the domain of definition of the function. The Gaussian process is {\it conditioned} on the given data $\mathcal{D}$ consisting of $f(x)$ measured at points $x$, so $f(x') \sim N\bigl(\mu(x'), k(x,x';\theta\bigr))$ where $N\bigl(\mu(x'), k(x,x';\theta)\bigr)$ is the Gaussian process with mean function $\mu(x')$ and $k(x,x';\theta)$ is the covariance function between the pair of points $x$ and $x'$. The covariance function between the pair of points depends on the hyperparameters $\theta$. A complete description of Gaussian process regression is provided in the book by~\citet{Rasmussen_GP_Book}.

In this work, the covariance function $k(x,x')$ (suppressing the vector $\theta$ of hyperparameters from the notation) used for Gaussian process regression is the Matern kernel which has the following form:
\begin{equation}
    k(x_i,x_j) = \frac{1}{\Gamma(\nu) 2^{\nu-1}} \left(\frac{\sqrt{2\nu}}{l_k} d(x_i,x_j) \right)^\nu K_\nu \left(\frac{\sqrt{2\nu}}{l_k}d(x_i,x_j) \right) \ ,
\end{equation}
where $k(x_i,x_j)$ is the covariance of the points $x_i$ and $x_j$, $\nu$ is a kernel parameter (hyperparameter), $\Gamma(\nu)$ is the gamma function evaluated at $\nu$, $l_k$ (hyperparameter) is the kernel length scale (or characteristic length scale of the Gaussian process), $d(x_i,x_j)$ is the Euclidean distance between $x_i$ and $x_j$, and $K_\nu (\ \cdot\ )$ is the modified Bessel function of order $\nu$. More details are available in the \texttt{scikit-learn} Matern kernel documentation~\citep{scikit-learn}. The critical hyperparameters of the Matern covariance function (kernel) are $\nu$ and $l_k$, which are discussed below.

The behavior of the Matern kernel depends on the value of the hyperparameter $\nu$. When $\nu={1}/{2}$, this kernel reduces to the absolute exponential kernel. When $\nu \to  \infty$, the Matern kernel reduces to the squared exponential kernel. For $\nu={3}/{2}$ and ${5}/{2}$, the associated Gaussian process is once- and twice-differentiable, respectively.

The length scale $l_k$ of the Matern kernel is a hyperparameter that needs to be optimized for a given dataset $\mathcal{D}$. In this work, we use \texttt{scikit-learn} for the Gaussian process regression, which automatically fits this hyperparameter to the available data. However, an initial guess for $l_k$ needs to be specified, and tuning this initial guess can lead to a more efficient fitting of the Gaussian process which, in turn, improves the subsequent Bayesian optimization performance~\citep{BayesOptPython}. We reparameterize this initial guess in terms of a relative length scale $l$. The $l_{k,c_i}$ length scale parameter associated with a model coefficient $c_i$ is determined as follows:
\begin{equation}\label{eq:lengthscale}
    l_{k,c_i} = l (c_{i,\text{upper}} - c_{i,\text{lower}}) \ ,
\end{equation}
where $l_{k,c_i}$ is an initial guess for the kernel length scale associated with coefficient $c_i$. Moreover, $ c_{i,\text{upper}}$ and $c_{i,\text{lower}}$ are the upper and lower bounds of the domain of definition for the coefficient $c_i$, respectively. Also, $l_k$ is the length scale vector with components $\bigl(l_{k,c_1},l_{k,c_2},...,l_{k,c_N}\bigr)$. This vector defines an anisotropic Matern covariance function (kernel) in the sense that the length scale associated with each model coefficient $c_i$ can have a different value. The motivation behind parameterizing $l_k$ as a vector is to specify a scale-appropriate initial guess for $l_k$ (viz., an appropriate scale for each model coefficient). For example, two model coefficients $c_1 \in [0.5,1.5]$ and $c_2 \in [0.001,0.005]$ will require two different length scales. Eq.~(\ref{eq:lengthscale}) will apply the same relative length scale to both dimensions, resulting in scale-appropriate length scales for each dimension. Finally, we note that the hyperparameter vector $\theta$ for the Matern covariance function (kernel) has the explicit form $\theta = (\nu,l_k^\text{T})^\text{T}$ where the superscript ``$\text{T}$'' denotes vector transposition.

\subsection{Utility functions}\label{sec:utility}
\subsubsection{Upper confidence bound (UCB) utility function}\label{sec:ucb}
The upper confidence bound (UCB) utility function features a clear exploration/exploitation tradeoff that is controlled by the $\kappa$ hyperparameter. The UCB utility function is given by
\begin{equation}
    \alpha_{\text{UCB}}(x) = \mu(x) + \kappa \sigma(x) \ ,
\end{equation}
where $\alpha_{\text{UCB}}(x)$ is the utility function value at point $x$, $\mu(x)$ is the expected value of the objective function, and $\sigma(x)$ is the standard deviation at point $x$ (viz., $\sigma(x) = k(x,x)^{1/2}$). Both $\mu(x)$ and $\sigma(x)$ are readily available from the Gaussian process regression of the objective function. Larger values of $\kappa$ lead to more exploration of points in the parameter space with large uncertainty, and smaller values of $\kappa$ lead to more exploitation of regions in the parameter space where the expected value $\mu(x)$ is known to be large.

\subsubsection{Probability of improvement (POI) utility function}\label{sec:poi}
The probability of improvement (POI) utility function returns the probability that a given point $x$ will result in an improvement of the objective function value. The POI hyperparameter $\xi$ specifies a minimum probability of improvement. The POI utility function is given by
\begin{equation} 
    \alpha_{\text{POI}}(x) = \text{CDF}\left(\frac{\mu(x) - \mu^\star -\xi}{\sigma(x)}\right) \ ,
\end{equation}
where $\alpha_{\text{POI}}(x)$ is the utility function value at point $x$, $\text{CDF}(z) $ is the cumulative distribution function of the standard Gaussian distribution (viz., a Gaussian distribution with zero mean and unit variance), $\mu(x)$ is the expected value of the objective function, $\mu^\star$ is the best objective function value observed so far, and $\sigma(x)$ is the standard deviation at $x$. Again, $\mu(x)$ and $\sigma(x)$ are immediately available from the Gaussian process regression of the objective function. Larger values of $\xi$ lead to a more explorative behavior, whereas smaller values of $\xi$ lead to a more exploitative behavior.

\subsubsection{Expected improvement (EI) utility function}\label{sec:ei}
The expected improvement (EI) utility function improves on the POI utility function. Whereas the POI utility function finds the point with the highest probability of providing any improvement, the EI utility function finds the point with the highest expected improvement. The hyperparameter $\xi$ is also used to control the exploration/exploitation tradeoff for the EI utility function. The EI utility function has the following form:
\begin{equation}
    \alpha_{\text{EI}}(x) = \text{CDF}\left(\frac{\mu(x) - \mu^\star - \xi}{\sigma(x)} \right) (\mu(x) - \mu^\star - \xi)  +  \text{PDF}\left(\frac{\mu(x) - \mu^\star - \xi}{\sigma(x)} \right)\sigma(x) \ ,
\end{equation}
where all variable definitions are the same as those of the POI function (Section~\ref{sec:poi}), and $\text{PDF}(z)$ is the probability density function for a standard Gaussian distribution. As in the case of the POI utility function, smaller $\xi$ values lead to more exploitation, whereas larger $\xi$ values lead to more exploration, in the parameter space.

\subsection{Generalized Error and Default Coefficient Preference (GEDCP) objective function}\label{sec:gedcp}
Objective function design is important for optimization. In general, the purpose of turbo-RANS is to reduce the error in a RANS prediction by adjusting the turbulence model coefficients. Therefore, the objective function should represent errors in the RANS predictions that ensure a well-posed optimization problem. In view of this, we formulate an objective function for use with turbo-RANS and demonstrate its utility.

We formulate a functional form for an objective function, which will be referred to henceforth as the generalized error and default coefficient preference (GEDCP) function. This function aims to be both highly versatile in terms of the types of reference data that can be used in the optimization and well-posed in terms of the process of the optimization itself. This function computes an error measure that can be used to assimilate dense reference data, sparse reference data, or integral reference data (or any combination of these types of data), making it adaptable generally for any kind of calibration problem. For example, RANS coefficients can be calibrated so that the predicted velocity field matches sparse wind-tunnel velocity data or so that the predicted force coefficients match those obtained from experimental data. For unsteady simulations, the coefficients can be calibrated based on an experimental Strouhal number or on a time-averaged force coefficient. The GEDCP function can also combine different types of reference data into a single objective function. For example, coefficients can be calibrated to simultaneously reduce errors in sparse reference data and in reference data for an integral parameter (e.g., lift or drag coefficient).

An optional term is included in the GEDCP function that embeds a preference for the ``default" values of the turbulence model coefficients. In practice, including this term serves two goals: namely, it embeds prior knowledge of the problem into the objective function, and it often results in the optimization problem having a single global optimum. For an example that motivates the first goal, consider the prior knowledge that $C_\mu = 0.09$ is commonly used in linear eddy viscosity models. This prior knowledge may be an important preference that is relevant for some applications. This default value is based on the correct ratio of $\nu_t \varepsilon/k^2$ in a turbulent channel flow away from the wall~\citep{Pope2000}. Here, $\nu_t$ is the eddy viscosity, $\varepsilon$ is the dissipation rate, and $k$ is the turbulent kinetic energy. If the optimized $C_\mu$ deviates from this value, there must be a corresponding significant reduction in the misfit error for this state of affairs to be acceptable. This preference penalty must also increase with increasing deviation from the default value of the coefficient, so if the nominally optimal value for $C_\mu$ is very far from 0.09, a large reduction in the fitting error to the reference data must occur for this to be acceptable. The second goal of the preference term is to ensure that the optimization problem has a single (unique) solution. In our experience, an error-based objective function based solely on the misfit between the model predictions and the reference data for calibrating RANS model coefficients often exhibits a saturation in the misfit. In other words, if the coefficient exceeds a certain value, the misfit error remains constant and, as a result, the optimization problem does not have a unique solution. Saturation or asymptotic behavior in the misfit can arise from the nature of the physics in the problem or from the presence of minimum, maximum, hyperbolic tangent, and other functions that are used in various combinations in many turbulence models. With a preference for the default value in a model coefficient, a small slope is introduced into the objective function, which drives coefficients towards the default value of the coefficient if this circumstance arises.

The GEDCP objective function, which depends on the model coefficients $\bigl\{c_1,c_2,\ldots,c_N\bigr\}$ that are to be calibrated against the available reference data, is formulated as follows:
\begin{align}
    f_\text{GEDCP}(c_1, c_2,\ldots,c_N) &=  -(\lambda_\text{F} E_\text{F} + \lambda_\text{I}E_\text{I})(1+\lambda_p p)\big|_{ c_1,c_2,\ldots,c_N}\ , \label{eq:gedcp_general} \\
    E_{\text{F}}(c_1, c_2,\ldots,c_N) &= \sum_{\phi \in \Phi} \text{MAPE}(\phi) \big|_{c_1, c_2,\ldots,c_N}= \sum_{\phi \in \Phi} \frac{1}{N_\phi}\sum_{i = 1}^{N_\phi} \left| \frac{\phi_i - \tilde \phi_i (c_1,c_2,...c_N)}{\phi_i} \right| \label{eq:gedcp_field},\\
    E_\text{I}(c_1,c_2,\ldots,c_N) &= \sum_{\psi \in \Psi} \left| \frac{\psi - \tilde \psi (c_1,c_2,\ldots,c_N) }{\psi}\right|\label{eq:gedcp_integral},\\
    p (c_1,c_2,\ldots,c_N)&= \frac{1}{N}\sum_{n=1}^{N} \left| \frac{c_{n,\text{default}} - c_n}{c_{n,\text{default}}}\right|\ .   \label{eq:gedcp_preference}
\end{align}
The variables and terms in Eqs.~(\ref{eq:gedcp_general})--(\ref{eq:gedcp_preference}) are defined below.

In Eq.~(\ref{eq:gedcp_general}), $f_\text{GEDCP}(c_1,c_2,...,c_N)$ is the objective function which depends on the tuneable coefficients $c_1, c_2, \ldots, c_N$; $E_\text{F}$ is the field error; $E_\text{I}$ is the integral parameter error; $p$ is the default coefficient preference term; and, $\lambda_\text{F}$, $ \lambda_\text{I}$, and $\lambda_p$ are regularization parameters that control the amount of preference (relative importance) ascribed to each term. The negative sign in Eq.~(\ref{eq:gedcp_general}) converts the problem of minimizing $E_\text{F}, E_\text{I}$, and $p$ to a maximization problem, which is the convention used for Bayesian optimization.

The reference data used for calibration of the model coefficients consist of a set of field values (e.g., velocity, pressure) and integral parameters (e.g., lift coefficient). The set of calibration fields is specified by $\Phi$, and the individual calibration fields are denoted by $\phi\in\Phi$. Individual reference values for a given calibration field are denoted by $\phi_i$. In a similar manner, the set of reference integral parameters is specified by $\Psi$, and the individual integral parameters are denoted by $\psi\in\Psi$. As an example, given calibration data consisting of 15 locations with pressure measurements $P$, 50 locations with measured velocity components $U$ and $V$, a reference lift coefficient $c_l$, and a reference drag coefficient $c_d$, we have the following sets: $\Phi = \bigl\lbrace P, U, V \bigr\rbrace$, $N_P=15$, $N_U=N_V=50$, and $\Psi = \bigl\lbrace c_l, c_d\bigr\rbrace$.

The field error $E_\text{F}$ [Eq.~(\ref{eq:gedcp_field})] is a summation of the mean absolute percentage errors (MAPE) in each field $\phi$ over the set of fields in the set $\Phi$ for which reference data are available for calibration. Here, $i=1,2\ldots,N_\phi$ indexes the points where reference data corresponding to the field $\phi$ are available. Furthermore, $\phi_i$ is the reference datum associated with the field $\phi$ at a given location and $\tilde \phi_i(c_1,c_2,\ldots,c_N)$ is the predicted value of the field at this location obtained using the set of model coefficients $\bigl\{c_1,c_2,\ldots,c_N\bigr\}$. This term can accommodate arbitrarily sparse and/or dense reference data and involves the determination of MAPE over all available reference data for a given field. Moreover, the use of MAPE implies that the influence of the reference data is not affected by the scale of $\phi_i$. For example, a 25$\%$ error at a point with low velocity in the boundary layer is penalized equally to that of a 25$\%$ error at a point with high velocity. The use of MAPE here is also motivated so that varying scales of $\phi\in\Phi$ are treated equally; for example, the velocity specified in units of m~s$^{-1}$ can have a different scale than that associated with the pressure in units of Pa. This equal treatment would not hold true if a mean-squared error (MSE) or mean-absolute error (MAE) were to be used in the formulation of the GEDCP objective function. 

The integral parameter error $E_\text{I}$ [Eq.~(\ref{eq:gedcp_integral})]  is a summation of the percentage error in each integral parameter for which reference data are available. Here, $\Psi$ is the set of integral parameters. Again, the percentage error is used so that varying scales associated with different quantities of interest within $\Psi$ are treated equally. For example, consider the calibration of an aerodynamic simulation to match reference data for the lift $c_l$ and drag $c_d$ coefficients. A given prediction using model coefficients $c_1,c_2,\ldots,c_N$ will provide an estimate for the lift $\tilde c_l(c_1,c_2,\ldots,c_N)$ and drag $\tilde c_d(c_1,c_2,\ldots,c_N)$ coefficients. The use of a relative error implies that even though $c_l$ and $c_d$ can have different orders of magnitude, the percentage errors will be penalized equally.

The default coefficient preference term $p$ is given by Eq.~(\ref{eq:gedcp_preference}). This term can be interpreted as the mean relative deviation of all model coefficients from their corresponding default values. The use of a relative (rather than an absolute) deviation is again motivated by possible scale differences in the values of the model coefficients.

In Eq.~(\ref{eq:gedcp_general}), $\lambda_\text{F}$, $\lambda_\text{I}$, and $\lambda_p$ can be adjusted to determine the preference for reference field data, reference integral data, and for default values of the model coefficients. We recommend setting $\lambda_\text{F}=\lambda_\text{I}=1$, and $\lambda_p=1/2$. These settings retain some preference for the default values of the model coefficients while ensuring that the objective function is primarily influenced by the error-based terms $E_\text{F}$ and $E_\text{I}$. If no reference field data is available (viz., $\Phi=\emptyset$), then $\lambda_\text{F}=0$. Similarly, the absence of reference integral parameter data  (viz., $\Psi=\emptyset$) implies $\lambda_\text{I}=0$.

The multiplicative design incorporated in Eq.~(\ref{eq:gedcp_general}) is to allow deviations from the default values of the model coefficients, as long as a corresponding significant decrease in misfit (prediction) error is realized. If the optimizer begins to ``stray'' too far from the default values of the model coefficients with only a marginal or negligible reduction in the prediction error, then the default coefficient preference term will penalize this behavior. However, if this departure from the default values of the model coefficients results in a significant prediction error reduction, then the objective function will still reward the proposed departure in the optimization process. For example, in the event that the proposed model coefficients result in perfect predictions of the quantities of interest (viz., $E_\text{F}, E_\text{I} \to 0$), then the default coefficient preference term has no effect. As long as the calibration of the model coefficients results in some error in the fitting process (which is expected for the majority of use cases), then the default coefficient preference term will influence the objective function.

As part of the hyperparameter optimization in this work (see Section~\ref{sec:hyperparameters}), the GEDCP loss functions for two example problems (see Sections~\ref{sec:airfoil}--~\ref{sec:phll}) were computed on a dense grid of RANS predictions associated with a grid of values of the model coefficients. This computation allows for the visualization of the GEDCP loss functions for two typical model coefficient calibration problems.

\begin{figure}
    \centering
    \includegraphics{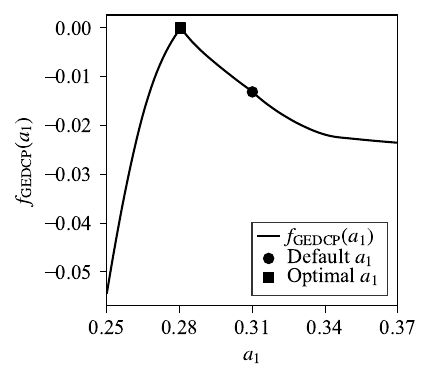}
    \caption{Objective function [Eq.~(\ref{eq:gedcp_airfoil})] for the airfoil case. The objective function is computed for a set of 50 RANS predictions with $a_1$ taking values in the interval $[0.25,0.37]$.}
    \label{fig:airfoil_obj_function}
\end{figure}

Fig.~\ref{fig:airfoil_obj_function} shows the GEDCP loss function associated with the calibration of the $a_1$ coefficient of the $k$-$\omega$ SST turbulence closure model. The calibration of this model coefficient was conducted in order to improve the predictive performance of the turbulence model for the lift coefficient of an airfoil. More details of this example problem are provided in Section~\ref{sec:airfoil}. Over a wide range of values of $a_1$, the GEDCP objective function for this case exhibits a single global optimum point and is a relatively straightforward function to optimize. For this example, turbo-RANS is able to find the optimum value of $a_1=0.28$ in approximately 9 evaluations of the objective function (see Section~\ref{sec:airfoil} for more detailed results). In the range of $0.34<a_1<0.37$, a small negative slope can be seen in $f_\text{GEDCP}(a_1)$. This small negative slope is due to the presence of the default coefficient preference term in the GEDCP objective function. Without the inclusion of this default coefficient preference term, an objective function that involves only a misfit term would have a zero slope here and, as a result, the optimizer would not be ``guided away'' from the upper bound for $a_1$ when exploring this plateau in the objective function.

\begin{figure}
    \centering
    \includegraphics{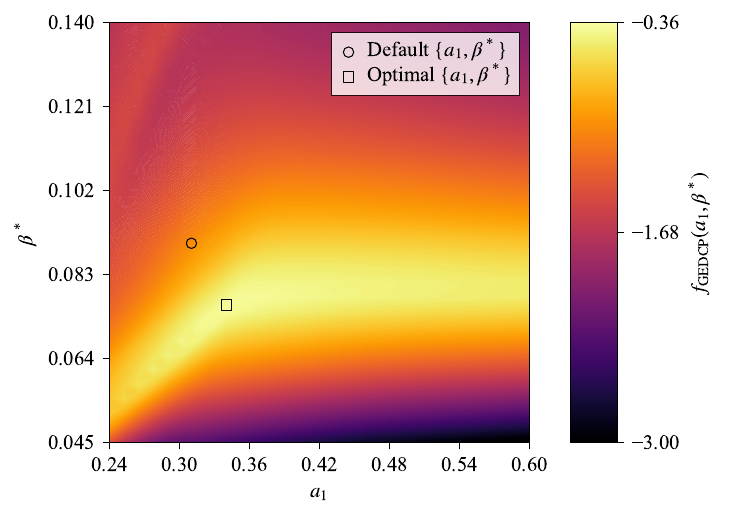}
    \caption{Objective function [Eq.~(\ref{eq:gedcp_phll})] for the periodic hills case. The objective function is computed from a set of 900 RANS predictions with $a_1 \in [0.24,0.60]$ and $\beta^*\in[0.045,0.140]$.}
    \label{fig:phll_obj_function}
\end{figure}

A more involved example that justifies the inclusion of the default coefficient preference term in the objective function is exhibited in Fig.~\ref{fig:phll_obj_function}. The GEDCP objective function shown here corresponds to optimizing the two model coefficients $a_1$ and $\beta^*$ of the $k$-$\omega$ SST turbulence closure model in order to improve the predictive performance of the model for a flow over periodic hills. The misfit error used in this example is a summation over the entire flow field because dense DNS data is available for this case. More details on this case are provided in Section~\ref{sec:phll}. Again, a single local maximum over a wide range of values of the model coefficients exists in this case. Without the inclusion of the default coefficient preference term, the objective function would asymptotically saturate (viz., achieve a maximum value) along a line defined by $\beta^*\approx 0.08$ in the range $0.36<a_1^*<0.60$. With a pure misfit error-based objective function, in spite of the fact that there is virtually no improvement in the predictive accuracy for increasing values of $a_1$, the optimizer in this case nevertheless recommends that the best estimate of $a_1=0.60$ (i.e., this value for $a_1$ maximizes the objective function). This large value for $a_1$ is not justified (or even reasonable) because there is only a marginal improvement in the predictive performance for values of $a_1$ greater than about 0.34. The GEDCP objective function addresses this problem through the inclusion of a small preference for the default values of the model coefficients. In so doing, a single (unique) solution is obtained for the optimization problem, resulting in reasonable values for the model coefficients: namely, $\beta^*=0.076$ and $a_1=0.34$. This optimum is found in approximately 20 evaluations of the objective function (see Section~\ref{sec:phll} for more results).

\subsection{turbo-RANS framework}\label{sec:algorithm}

\begin{figure}
    \centering
    \includegraphics{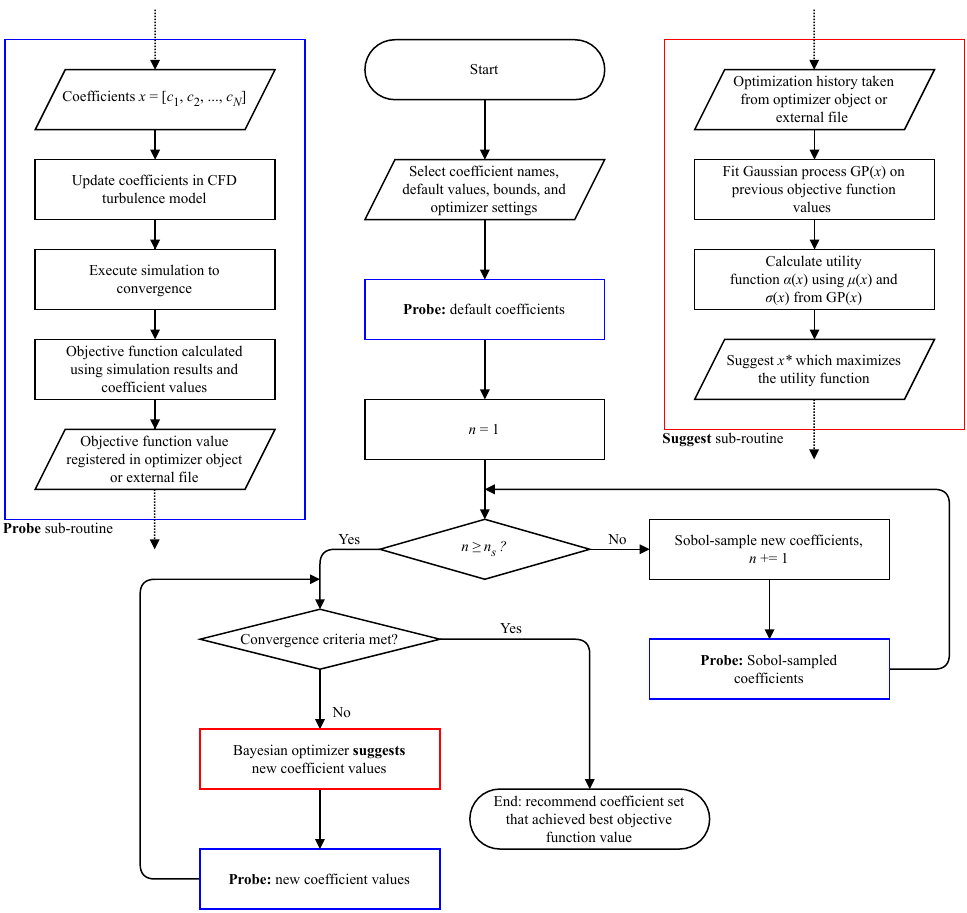}
    \caption{The turbo-RANS framework summarized in an algorithm flowchart.}
    \label{fig:turbo-RANS_algorithm}
\end{figure}

Fig.~\ref{fig:turbo-RANS_algorithm} shows the turbo-RANS framework, along with details of the probe-and-suggest sub-routines. Various steps of the algorithms incorporated in this framework will be described below.

To initialize the optimization process, information must be provided by the user as to which model coefficients are to be optimized. This information includes the default values for the model coefficients and their lower and upper bounds. Hyperparameters (see Section~\ref{sec:hyperparameters}) used in the optimization, as well as other settings, must also be specified by the user as part of the initialization. It is noted that the hyperparameters can be modified during each iteration of the optimization process. The first point probed is the point corresponding to the default values for the model coefficients. If the chosen optimization problem does not include default values for the model coefficients, then this first default coefficient probe is skipped in the initialization.

After probing the default values for the model coefficients, the sampling loop is executed. In the framework, $n_s$ is the number of sample points that are probed before the Bayesian optimization loop is entered. If $n_s$ is too small, errant behavior of the Gaussian process regression can occur (see Section~\ref{sec:hyperparameters}). At this stage, values of the model coefficients are sampled using Sobol sampling to produce a Sobol sequence of points, which promotes better uniformity of the sampled values compared to a purely random sampling~\citep{Sobol1967}. After a new point is obtained using Sobol sampling, this point is probed. The sampling loop continues until a number of points $n_s$ (prespecified by the user) have been probed.

The Bayesian optimization loop begins after a default point and a sufficient number of additional sample points have been probed. In the Bayesian optimization loop, the optimizer considers the past history and suggests a new value of the model coefficients (viz., a point) to be probed. Following this, the suggested new point is probed. The probe-suggest loop continues until a convergence criterion is satisfied. This convergence criterion is specified by the user. We envision that turbo-RANS loops that involve relatively inexpensive CFD calculations can be stopped when the best objective function value and/or the best set of model coefficients do not provide any further improvements (changes) in the optimization process after a certain number of iterations (early stopping). For problems corresponding to more expensive CFD calculations, an upper bound on the computational budget can be prescribed {\it a priori\/} by the user (e.g., repeat the loop until a pre-specified number of CPU hours has been expended in the computation). With either the convergence-based or budget-based stopping criteria, the Bayesian optimization algorithm will result in an efficient optimization in terms of the number of objective function evaluations (viz., CFD calculations). After the stopping criteria are satisfied, the final recommended model coefficient set is the set that achieves the maximum value of the objective function.

Probing a set of values of the model coefficients consists of running a RANS calculation with these coefficients, computing the objective function, and registering the objective function value so that it can be used in future iterations. The probe sub-routine in Fig.~\ref{fig:turbo-RANS_algorithm} details these steps. The convergence criteria for the RANS calculations are also specified (prescribed) by the user. We recommend the standard practice of evaluating both the residual values of various flow quantities and monitoring an integral parameter to assess convergence. In terms of actual implementation, this process is typically automated using scripts that modify the turbulence model coefficients, execute a RANS calculation, and monitor the convergence of this calculation. To reduce computational demands, we recommend using the converged flow fields obtained from the default coefficient probe (viz., the default values of the model coefficients) as initial conditions for future probe RANS calculations. This re-use of the converged flow field from the default coefficient probe greatly reduces the computational costs of the turbo-RANS algorithm, since the flow fields start near convergence. The post-processing of RANS calculations into fields that are compatible with the GEDCP objective function (Section~\ref{sec:gedcp}) is also typically automated. For example, this post-processing can include computing force coefficients or sampling the solution field at the locations where the reference data is available. Details of the objective function registration are provided in Section~\ref{sec:implementation}, and in the turbo-RANS documentation~\citep{turborans_github}.

The suggest sub-routine exhibited in Fig.~\ref{fig:turbo-RANS_algorithm} addresses two sub-optimization problems: namely, (1) fitting a Gaussian process using all previously probed points obtained in modelling of the objective function; and, (2) maximizing the utility function. These optimization problems are already implemented in the dependencies for turbo-RANS. The suggest sub-routine executes almost instantaneously, and therefore the main computational expense in the turbo-RANS algorithm is conducting a RANS calculation with a given set of turbulence model coefficients. The computational time of the suggest sub-routine increases with the number of points previously sampled, as the cost of the Gaussian process regression increases with an increasing number of points. However, in our experience, the computational cost of the suggest sub-routine is negligible compared to the cost of conducting the RANS calculations required for each point probed (undertaken in the probe sub-routine). For large parameter spaces and a large number of previously evaluated points defining the objective function hypersurface (e.g., greater than 100), the Gaussian process regression may perform slowly. However, even in the worst case, we have observed that the cost of the suggest sub-routine remains below 1$\%$ of the total cost associated with the RANS calculations required for each probed point.

\subsection{Implementation}\label{sec:implementation}
The core Bayesian optimization package used in turbo-RANS is the Python library \texttt{bayesian-optimization}~\citep{BayesOptPython}. This library is widely used in many Bayesian optimization applications and includes implementations of the UCB, POI, and EI utility functions. The underlying Gaussian process regression library used in \texttt{bayesian-optimization} is from \texttt{scikit-learn}~\citep{scikit-learn}. A series of convenient Python interfaces to these two libraries has been provided in the turbo-RANS GitHub repository~\citep{turborans_github}. The repository includes examples and templates for a wide variety of use cases, including the following:
\begin{itemize}
    \item two examples based on OpenFOAM: flow over an airfoil (Section~\ref{sec:airfoil}), and flow over periodic hills (Section~\ref{sec:phll});
    \item one example based on Ansys Fluent: flow through a converging-diverging channel (Section~\ref{sec:cndv});
    \item an example script for STAR-CCM+ users;
    \item templates for optimizing OpenFOAM simulations;
    \item solver-agnostic templates for optimizing CFD simulation Python scripts; and,
    \item solver-agnostic templates for optimizing CFD simulation shell scripts.
\end{itemize}
Full documentation for the code is provided in the GitHub repository~\citep{turborans_github}. It should be noted that the parallelization of Bayesian optimization is non-trivial. While a RANS calculation with a given set of model coefficients is typically parallelized, the primary turbo-RANS algorithm is currently restricted to running in a sequential manner. This restriction means that only a single RANS calculation is conducted at a time, rather than probing multiple sets of model coefficients to enable multiple RANS calculations to be undertaken simultaneously. It is assumed that the RANS calculation step is by far the most computationally expensive component of the turbo-RANS algorithm since the acquisition of a new point in the Bayesian optimization occurs nearly instantaneously. If desired, the user can opt to implement a parallelized Bayesian optimization by modifying the turbo-RANS example scripts. The code's verbose file input and output structure facilitates implementing parallelized Bayesian optimization but, for now, this functionality is left for a future effort.

\section{Demonstrations}\label{sec:demonstrations}
\subsection{Example 1: Flow over an airfoil}\label{sec:airfoil}
\subsubsection{Description}
RANS calculations are often used in external aerodynamics, such as in aerospace, wind engineering, and automotive engineering. Accurate prediction of integral quantities such as force coefficients is typically desired in these applications. The purpose of this example is to demonstrate how turbo-RANS can be used to calibrate turbulence model coefficients using integral parameter reference data. 

\subsubsection{Computational setup}\label{sec:airfoil_setup}
We calculate the steady-state, incompressible, turbulent flow over a NACA 0012 airfoil at a $10.12^\circ$ angle of attack and a chord-based Reynolds number of $Re_c=6 \times 10^6$. The case details are taken from the NASA turbulence modelling resource ``2D NACA 0012 Airfoil Validation" 2DN00 case~\citep{NASAturbmodel}. OpenFOAM v2206 was used---the RANS calculations were conducted using the semi-implicit method for pressure-linked equations (SIMPLE) solver (simpleFoam) in OpenFOAM. A second-order upwind scheme was used for the discretization of the convective terms in the momentum equation, and the central difference scheme was used for the discretization of the diffusion terms. A first-order upwind scheme was used for the discretization of the convective terms in the turbulence transport equations. The geometric agglomerated algebraic multigrid (GAMG) solver was used to obtain the pressure, and the preconditioned bi-conjugate gradient (PBiCGStab) solver was used to solve all the other equations. The full OpenFOAM case setup is available in the turbo-RANS GitHub repository~\citep{turborans_github}.

The turbulence model used for this case is the $k$-$\omega$ shear stress transport (SST) model. The OpenFOAM v2206 variant of this model is the updated 2003 Menter version~\citep{Menter2003}, with production terms from ~\citet{Menter2001} \citep{Openfoam}. The following model equations are provided in ~\citet{Menter2003} ($k$ is the turbulent kinetic energy and $\omega$ is the specific dissipation rate):
\begin{align}
    \frac{\partial(\rho k)}{\partial t} + \frac{\partial(\rho U_i k)}{\partial x_i} &= \tilde P_k - \beta^*\rho k \omega +\frac{\partial}{\partial x_i}\left[ (\mu + \sigma_k \mu_t) \frac{\partial k}{\partial x_i}\right]\ , \label{eq:k}\\
    \frac{\partial(\rho \omega)}{\partial t} + \frac{\partial (\rho U_i \omega)}{\partial x_i} &= \alpha \frac{\tilde P_k}{\nu_t} - \beta \rho \omega^2 + \frac{\partial}{\partial x_i}\left[ (\mu + \sigma_\omega \mu_t) \frac{\partial \omega}{\partial x_i}\right] + 2(1-F_1)\rho \sigma_{\omega 2} \frac{1}{\omega} \frac{\partial k }{\partial x_i} \frac{\partial \omega}{\partial x_i}\ , \label{eq:omega} \\
    \nu_t &= \frac{a_1 k }{\text{max} (a_1 \omega, SF_2)}\ ,\label{eq:nu_t}
\end{align}
where the definitions of the blending functions $F_1$ and $F_2$, and limited production term $\tilde P_k$ have been omitted as they are not relevant to the present discussion. Given that the coefficients are a blend of the coefficients of the $k$-$\varepsilon$ and $k$-$\omega$ turbulence closure models, the $k$-$\omega$ SST model has 10 model coefficients: $a_1$, $\beta^*$, $\sigma_{k1}$, $\sigma_{k2}$, $\alpha_1$, $\alpha_2$, $\beta_1$, $\beta_2$, $\sigma_{\omega1}$, and $\sigma_{\omega2}$. Default values for these coefficients are given as: $a_1 = 0.31$, $\beta^* = 0.09$, $\sigma_{k1} = 0.85$, $\sigma_{k2} = 1$, $\alpha_1 = 5/9$, $\alpha_2 = 0.44$, $\beta_1 = 3/40$, $\beta_2=0.0828$,  $\sigma_{\omega1} = 0.5$, and $\sigma_{\omega2}=0.856$. Here, the subscript 1 denotes the $k$-$\omega$ model, and the subscript 2 denotes the $k$-$\varepsilon$ model. The default values of these coefficients arise from the application of various heuristics, such as tuning the behavior of certain terms to be in conformance with the characteristics of specific types of (usually canonical) flows. For many industrially relevant flows, additional tuning of these model coefficients can increase predictive accuracy. For the airfoil case, we employ turbo-RANS to calibrate the $a_1$ coefficient of the $k$-$\omega$ SST turbulence model for this particular flow.

\begin{figure}
     \centering
     \begin{subfigure}[t]{0.33\textwidth}
         \centering
         \includegraphics{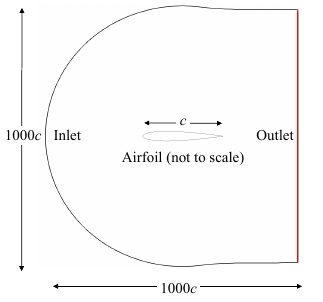}
         \caption{Computational domain and boundary names.}\label{fig:domain_airfoil}
     \end{subfigure}
     \begin{subfigure}[t]{0.33\textwidth}
         \centering
         \includegraphics{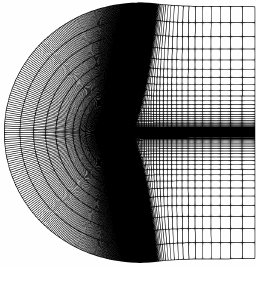}
         \caption{Computational mesh.}\label{fig:mesh_airfoil}
     \end{subfigure}
     \begin{subfigure}[t]{0.33\textwidth}
         \centering
         \includegraphics{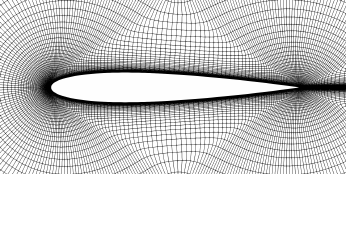}
         \caption{Close-up of computational mesh on the airfoil surface.}
     \end{subfigure}
        \caption{Computational setup for the airfoil example case.}
        \label{fig:setup_airfoil}
\end{figure}

Fig.~\ref{fig:setup_airfoil} shows the computational domain and mesh, which were taken from the NASA turbulence modelling resource~\citep{NASAturbmodel}. The computational domain is relatively large. More specifically, the domain is $500c$ ($c$ is the chord length) in any direction from the airfoil so as to negate blockage effects. Pressure at the inlet is zero gradient, and all other flow quantities at the inlet are prescribed as follows:
\begin{align}
    U_\text{inlet} &= (50.68, 0, 9.05)  \ \text{m}\text{ s}^{-1}\ ,\label{eq:U_inlet}\\
    k_\text{inlet} &= 0.001075 \  \text{m}^2\text{ s}^{-2}\ , \\
    \omega_\text{inlet} &= 0.3279 \ \text{s}^{-1}\ .\label{eq:omega_inlet}
\end{align}
In particular, turbulence quantities at the inlet are based on a 0.052\% turbulence intensity and a turbulent length scale of $10\%$ of the chord length. At the outlet plane of the computational domain, all quantities except pressure are zero-normal-gradient, and the gauge pressure is fixed at zero Pa.

The mesh for this case consists of 57,344 structured hexahedral cells, with approximately 129 points along the surface of the airfoil. With the default values for $k$-$\omega$ SST model coefficients, the $y^+$ wall-normal distance from the airfoil surface has values lying between 0.017 and 0.88, with a mean value of 0.30. These values imply that the RANS computation for this case is wall-resolved.

\subsubsection{Optimization problem}

The purpose of the airfoil example case is to optimize the $a_1$ model coefficient in the $k$-$\omega$ SST turbulence model in order to improve the accuracy of the model for prediction of the lift coefficient $c_l$. The objective function for the airfoil case is given by
\begin{align}\label{eq:gedcp_airfoil}
    f_\text{GEDCP}(a_1) &= -(E_\text{I})(1+\tfrac{1}{2}p)\big|_{a_1}\ ,\\
    E_\text{I}(a_1) &= \left| \frac{c_l - \tilde c_l(a_1)}{c_l} \right|\ .
\end{align}
Here, $f_\text{GEDCP}(a_1)$ is the GEDCP objective function [Eq.~(\ref{eq:gedcp_general})], with $\lambda_\text{I} = 1$, and $\lambda_p=1/2$. This objective function primarily aims to minimize the relative error in the prediction of the integral parameter $c_l$ while retaining a minor preference for the default value of the model coefficient: namely, $a_1 = 0.31$. In the present work, we have computed this objective function over a grid of values of $a_1$ so that it can be visualized (cf.~Fig.~\ref{fig:airfoil_obj_function}). No field quantities are used in the objective function for the airfoil case. The parameter space for the model coefficient for the airfoil case is $a_1 \in [0.25,0.37]$. The coefficient $a_1$ was calibrated so that the predictions matched the reference lift coefficient of $c_l= 1.0707$, the latter of which comes from the fully-turbulent experimental measurement for this quantity obtained by ~\citet{Ladson1988} at a Mach number of 0.15, where the suction and pressure side transition points were fixed at a distance of $0.05 c$ from the leading edge, using No.~80-W grit carborundum strips.

\subsubsection{Results}
\begin{figure}
    \centering
    \includegraphics{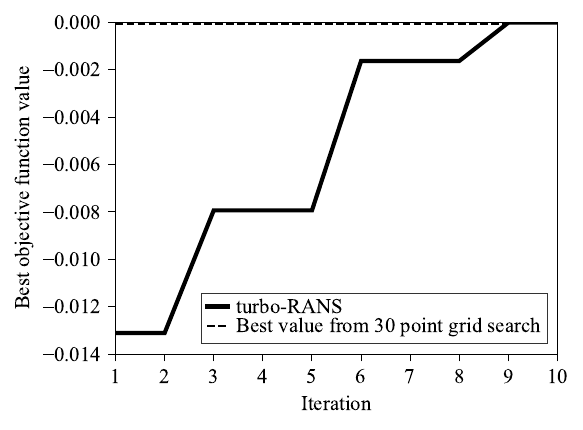}
    \caption{Convergence history for the optimization process for the airfoil example case.}
    \label{fig:airfoil_convergence}
\end{figure}

The turbo-RANS framework was applied to the airfoil example case, allowing a maximum of ten iterations in the optimization process. The first iteration was the default coefficient run (viz., RANS calculation using the default value for $a_1$). Subsequently, four additional Sobol-sampled iterations were completed. Therefore, five of these ten iterations correspond to iterations of the Bayesian optimization procedure. Each iteration corresponds to a RANS calculation with a given value for the $a_1$ model coefficient. The computational time was greatly reduced by using the converged fields from the default coefficient run as initial fields for the remaining RANS calculations involving the new proposals for the value of $a_1$. Fig.~\ref{fig:airfoil_convergence} displays the convergence history over the ten iterations.

In nine iterations, the optimization process attained a maximum value of the objective function, and the value of $a_1$ corresponding to this maximum (optimum) objective function value yielded a predicted value for the lift coefficient of $\tilde c_l = 1.0706$. This predicted value for the lift coefficient is in excellent conformance with the experimental value for the lift coefficient of $c_l=1.0707$. The airfoil example calibration results are summarized in Table~\ref{tbl:airfoil_summary}. While the default coefficient performs reasonably well at predicting the experimental value for the lift coefficient, this example demonstrated that turbo-RANS can efficiently calibrate the model coefficient so that the predicted value of the lift coefficient provides an essentially perfect agreement with the corresponding experimental measurement of this coefficient.

\begin{table}[]\caption{Summary of the airfoil lift coefficient calibration case.}\label{tbl:airfoil_summary}
\centering
\begin{tabular}{ccc}
\hline
Case                                                              & $a_1$ & $c_l$  \\ \hline
Experimental calibration data~\citep{Ladson1988} & ---    & 1.0707 \\
$k$-$\omega$ SST, default coefficient                             & 0.31  & 1.0847 \\
$k$-$\omega$ SST, turbo-RANS optimized                            & 0.28  & 1.0706 \\ \hline
\end{tabular}
\end{table}

\subsection{Example 2: Periodic hills}\label{sec:phll}

\subsubsection{Description}
Predicting the evolution of separated, turbulent flow is a major challenge for RANS turbulence models~\citep{Wilcox1994}. Non-equilibrium effects, non-local effects, and history effects are frequently present for this type of flow. Unfortunately, commonly used RANS turbulence models such as the $k$-$\omega$ SST model often do not accurately predict these flow phenomena. Nevertheless, these models are frequently applied to separated turbulent flows since separation occurs in many industrial flows. A canonical example of separation is turbulent flow over periodic hills. In this flow, a periodic domain is used so that the flow repeatedly separates from the smooth surface of one hill, reattaches, accelerates up a slope of the hill immediately downstream of it, and then separates again from the surface of this second hill. When we apply single-point RANS turbulence closure models such as the $k$-$\omega$ SST model to this type of flow, we can calibrate the model coefficients so that the effects of the poorly captured physics are at least mimicked. Otherwise, the turbulence model will be unable to accurately predict the general characteristics of the flow. In this example, we demonstrate how turbo-RANS can be used to calibrate turbulence model coefficients so that the predicted field quantities are in better conformance with the associated reference data.

\subsubsection{Computational setup}
We calculate the steady-state, incompressible, turbulent flow over a series of periodic hills at a height-based Reynolds number of $Re_H=5,600$. Direct numerical simulation data and OpenFOAM meshes are provided for this flow by ~\citet{Xiao2020}. The software, solver, and schemes are identical to the airfoil example case (cf.~Section~\ref{sec:airfoil_setup}). The $k$-$\omega$ SST turbulence model is also calibrated in this example [Eqs.~(\ref{eq:k})--(\ref{eq:nu_t})].

\begin{figure}
     \centering
     \begin{subfigure}[t]{0.4\textwidth}
         \centering
         \includegraphics{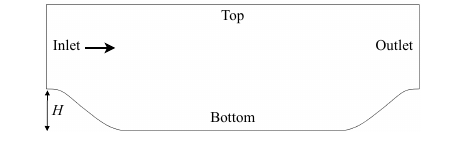}
         \caption{Computational domain and boundary names.}\label{fig:domain_phll}
     \end{subfigure}
     \begin{subfigure}[t]{0.4\textwidth}
         \centering
         \includegraphics{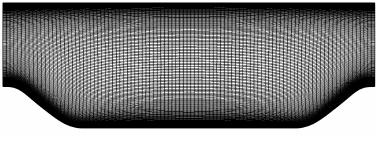}
         \caption{Computational mesh.}\label{fig:mesh_phll}
     \end{subfigure}
        \caption{Computational setup for the periodic hills example case.}
        \label{fig:setup_phll}
\end{figure}

The computational domain and mesh for this case are taken from Xiao et al.~\citep{Xiao2020}. The computational domain is two-dimensional (2D)---the three-dimensional (3D) DNS data has been averaged along the spanwise direction. The mesh consists of 14,751 hexahedral cells. Boundary conditions and mesh convergence for the RANS calculation, in this case, are discussed in~\citet{McConkeySciDataPaper2021}. In short, periodic boundary conditions are imposed at the inlet and outlet planes of the computational domain. A no-slip wall boundary condition is imposed on the top and bottom walls of the domain. Along these no-slip walls, $0.01 \leq y^+ \leq 0.80$, with an average value for $y^+$ along the top and bottom walls of $0.71$ and $0.20$, respectively. This range of values for $y^+$ along the top and bottom walls implies that the RANS calculations conducted here are wall-resolved. Figure~\ref{fig:setup_phll} shows the computational domain and mesh for the periodic hills case.

\subsubsection{Optimization problem}
For the case of the periodic hills, the GEDCP objective function is used, but flow field quantities are used in the misfit error term of the function. The purpose is to calibrate the $a_1$ and $\beta^*$ model coefficients of the $k$-$\omega$ SST turbulence model so that the predictions of the magnitude of the velocity $\vec{U}$ and $k$ fields are in better conformance with the available DNS reference data. No integral parameters are included in the objective function for this example. In consequence, the objective function assumes the following form:
\begin{align}\label{eq:gedcp_phll}
    f_\text{GEDCP}(a_1,\beta^*) &= -(E_\text{F})(1+\tfrac{1}{2}p)\big|_{a_1,\beta^*}\ ,\\
    E_{\text{F}}(a_1,\beta^*) &= \sum_{\phi \in \lbrace|\vec{U}|,k \rbrace} \text{MAPE}(\phi) \nonumber\\&= \sum_{\phi \in \lbrace|\vec{U}|,k \rbrace} \frac{1}{N_\phi}\sum_{i = 1}^{N_\phi} \left| \frac{\phi_i - \tilde \phi_i (a_1,\beta^*)}{\phi_i} \right| \ .\\
\end{align}
This objective function encapsulates a tradeoff between reducing the MAPE in the prediction of $|\vec{U}|$ and $k$ while retaining a minor preference for the default value of the model coefficients: namely, $a_1 = 0.31$ and $\beta^* = 0.09$. The parameter space corresponding to the model coefficients for the case of the periodic hills is $a_1 \in [0.24,0.60]$ and $\beta^* \in [0.045,0.14]$. The DNS reference data has been interpolated onto the RANS mesh so that $N_\phi=14,751$. Given that every point in the RANS calculation has corresponding reference data, this example demonstrates how turbo-RANS can be used in an application involving highly dense reference data. 

\subsubsection{Results}
A fixed computational budget of 30 iterations was prescribed for the optimization. The default values of the model coefficients were used for the first iteration. For the next nine iterations, Sobol sampling was used to generate the nine additional points. Subsequently, Bayesian optimization was used for the remaining 20 iterations.

\begin{figure}
    \centering
    \includegraphics{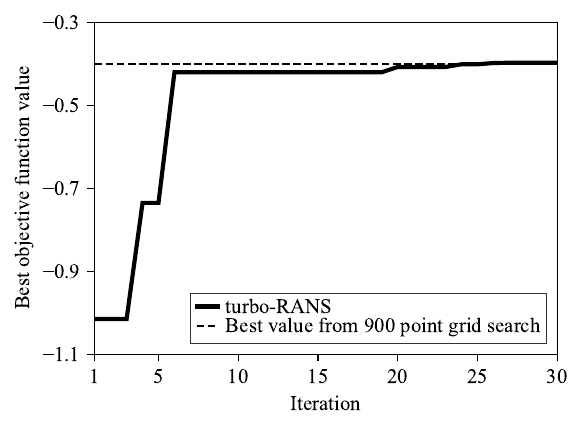}
    \caption{Convergence of the optimization process for the case of the periodic hills.}
    \label{fig:phll_convergence}
\end{figure}

Fig.~\ref{fig:phll_convergence} illustrates the convergence of the optimization process for the periodic hills example case. After approximately 20 iterations, the Bayesian optimization has found the maximum (optimum) value (dashed line corresponding to the maximum value obtained using a $30\times 30$ grid search in the parameter space). At the end of the optimization process, Bayesian optimization found model coefficients whose associated objective function value was slightly larger (better score) than the maximum (best score) obtained in the grid search---naturally, this optimal value occurs between the grid points used in the exhaustive grid search.

\begin{figure}
    \centering
    \includegraphics{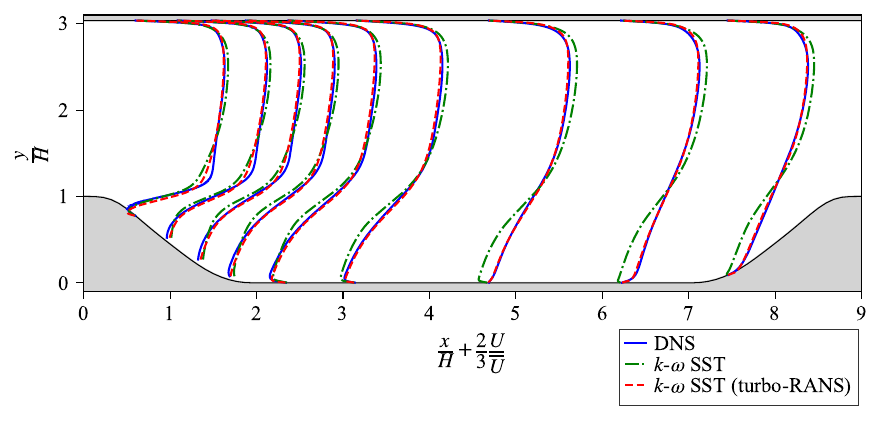}
    \caption{The streamwise ($U$) component of velocity along various vertical sampling lines for the case of the periodic hills.}
    \label{fig:phll_U_samples}
\end{figure}

\begin{figure}
    \centering
    \includegraphics{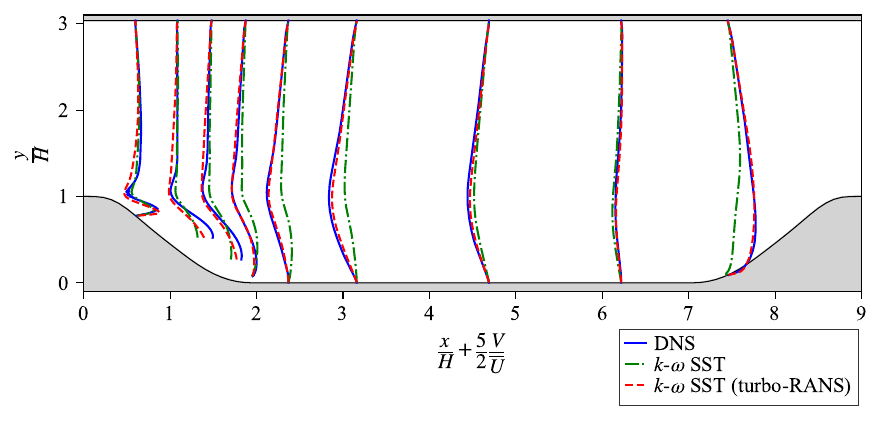}
    \caption{The vertical ($V$) component of velocity along various vertical sampling lines for the case of the periodic hills.}
    \label{fig:phll_V_samples}
\end{figure}

The predicted velocity and turbulent kinetic energy fields obtained using RANS calculations based on the default and optimized model coefficients were compared to the DNS reference data. Figs.~\ref{fig:phll_U_samples} and~\ref{fig:phll_V_samples} display the streamwise $U$ and vertical $V$ components of the velocity along several vertical sampling lines in the computational domain, respectively. The RANS predictions obtained using the turbo-RANS-optimized model coefficients agree very well with the DNS data, demonstrating a significant improvement in comparison to the corresponding RANS predictions obtained using the default values for the model coefficients. The default values of the model coefficients lead to predictions that overestimate the flow separation and, as a result, do not accurately predict the velocity after the occurrence of the separation. In the bulk flow over the separated flow region, the velocity predicted by the RANS models using the default values of the model coefficients is also high compared to the DNS reference data. After calibrating the $a_1$ and $\beta^*$ model coefficients, the velocity here agrees well with the DNS reference data. Both the default and optimized RANS predictions exhibit discrepancies in the near-wall quantities directly along the separated bump. Here, the breakdown of assumptions made in modelling the RANS boundary layers is likely responsible for this discrepancy. Overall, the inclusion of the DNS $U$ and $V$ reference data through the GEDCP objective function [Eq.~(\ref{eq:gedcp_phll})] leads to a greatly improved conformance between the RANS predictions and the DNS reference data for this separated flow, as summarized in Table~\ref{tbl:phll_mape}.

\begin{table}[]
\centering
\caption{The values of MAPE (lower is better) before and after turbo-RANS calibration of the model coefficients in the $k$-$\omega$ SST turbulence model for flow over periodic hills.}
\label{tbl:phll_mape}
\begin{tabular}{ccc}
\hline
                               & \multicolumn{2}{c}{Mean absolute percentage error (MAPE)}             \\
Coefficients                   & $k$-$\omega$ SST, default coefficients & $k$-$\omega$ SST, turbo-RANS \\ \hline
$\lbrace a_1, \beta^* \rbrace$ & $\lbrace 0.31, 0.09\rbrace$            & $\lbrace 0.34, 0.076\rbrace$ \\
$|\vec{U}|$                    & 0.522                                  & 0.120                        \\
$k$                            & 1.285                                  & 0.850                        \\ \hline
\end{tabular}
\end{table}

\begin{figure}
    \centering
    \includegraphics{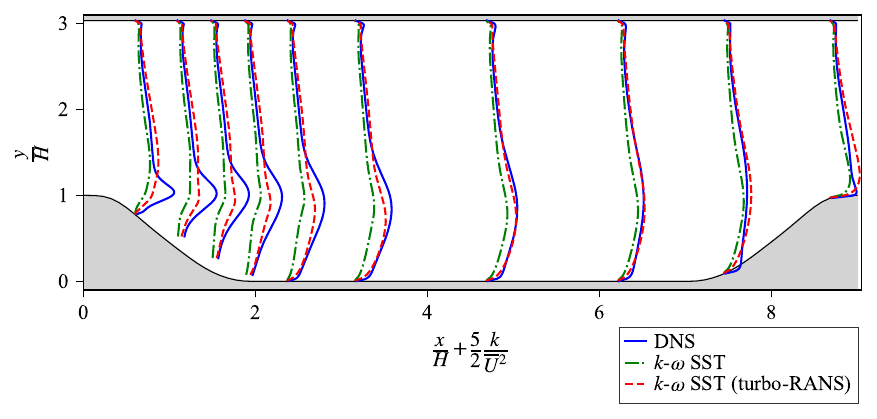}
    \caption{The turbulent kinetic energy $k$ along various vertical sampling lines for the case of the periodic hills.}
    \label{fig:phll_k_samples}
\end{figure}

Fig.~\ref{fig:phll_k_samples} compares vertical profiles of the turbulent kinetic energy $k$ along various sampling lines in the computational domain. The $k$ field was used in conjunction with the $|\vec{U}|$ field in the calibration of the model coefficients. While there is a clear improvement in the $k$ estimation in many regions of the flow, there is still a discrepancy between the DNS reference $k$ and the RANS predicted $k$ in the shear zone at the top of the separation region. Further accuracy improvements may be possible through the inclusion of additional model coefficients in the calibration procedure, such as the various coefficients that appear in the $k$-transport equation [Eq.~(\ref{eq:k})] and the $\omega$-transport equation [Eq.~(\ref{eq:omega})]. Nevertheless, the calibration procedure has generally resulted in an improved prediction of $k$ for this separated flow, as summarized in Table~\ref{tbl:phll_mape}.

\begin{figure}
    \centering
    \includegraphics{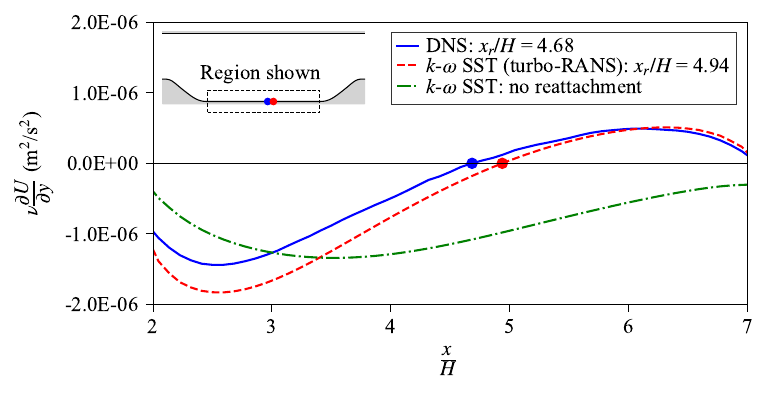}
    \caption{The wall shear stress along the bottom wall for the case of the periodic hills.}
    \label{fig:phll_WSS}
\end{figure}

While the reference data used in the GEDCP objective function was $|\vec{U}|$ and $k$, it was of interest to determine whether other flow quantities were also predicted more accurately. Even though the wall shear stress was not included as an explicit calibration target, Fig.~\ref{fig:phll_WSS} shows that the wall shear stress along the bottom wall is in much better conformance with the DNS data of the wall shear stress (not used in the calibration) after calibration of the model coefficients. In fact, the baseline $k$-$\omega$ SST turbulence model does not predict any reattachment of the flow along the bottom wall before the flow reaches the downstream hill. After calibration, the $k$-$\omega$ SST turbulence model now predicts reattachment, with the reattachment point in very good agreement with the DNS data. Fig.~\ref{fig:phll_WSS} demonstrates that calibrating turbulence model coefficients based on a specific set of flow quantities can lead to improvements in the predictive accuracy of other flow quantities not used in the calibration.

\subsection{Example 3: Converging-diverging channel}\label{sec:cndv}

\begin{figure}
    \centering
    \includegraphics[width = 13 cm]{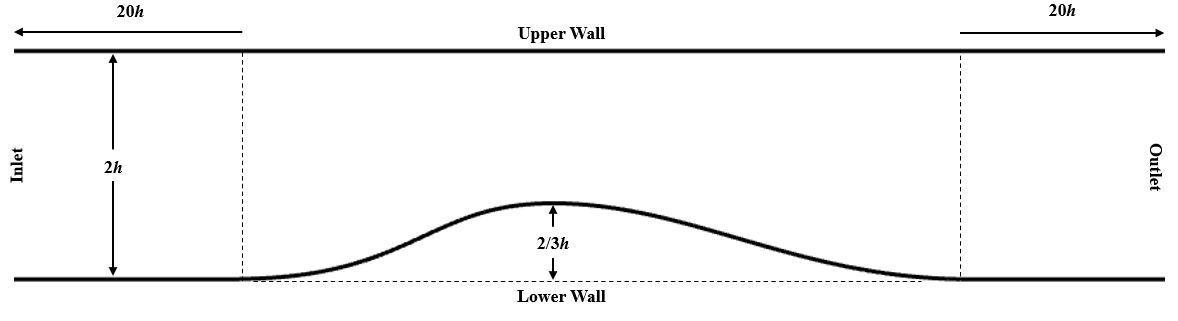}
    \caption{Computational domain for the converging-diverging channel.}
    \label{fig:domain_channel}
\end{figure}

\begin{figure}
    \centering
    \includegraphics [width = 10 cm]{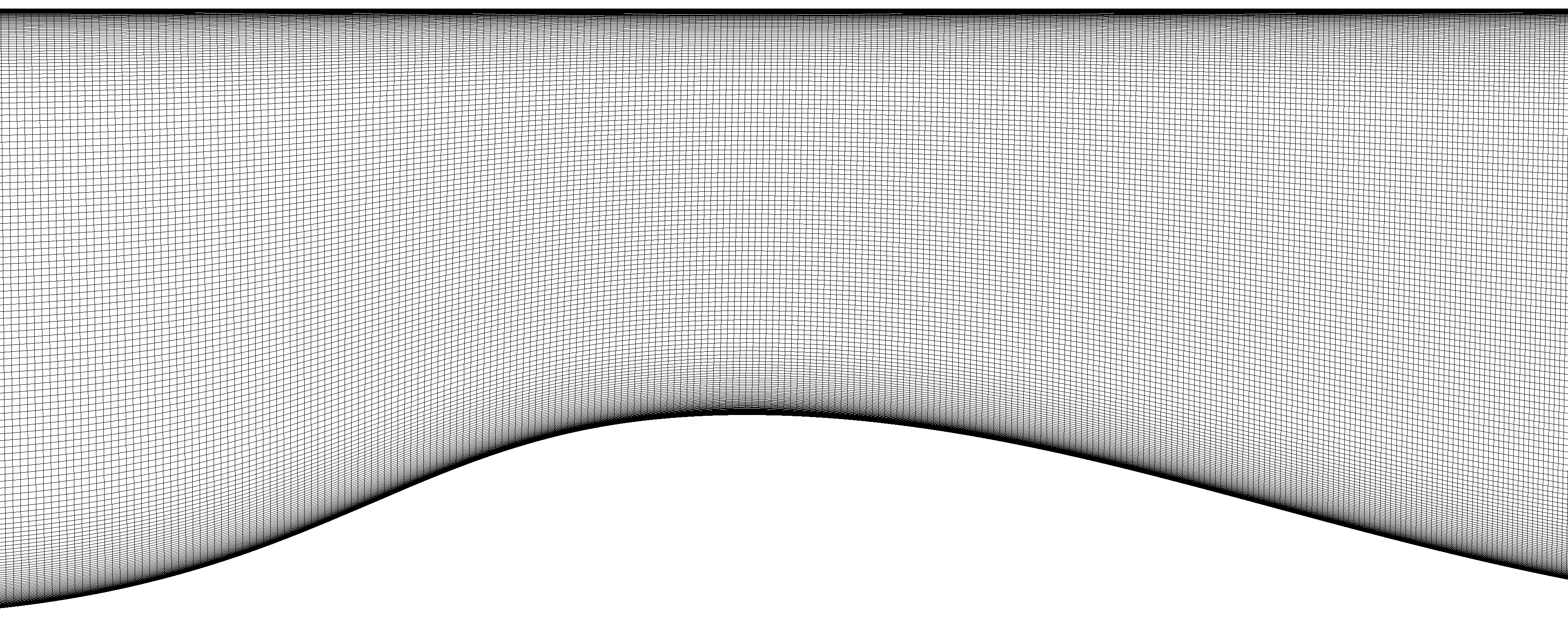}
    \caption{Mesh for the converging-diverging channel~\citep{McConkeySciDataPaper2021}.} 
    \label{fig:mesh_channel}
\end{figure}

\subsubsection{GEKO turbulence model}\label{sec:cndv_setup}
The latest turbulence model that has been developed by Ansys, Inc.~is the ``generalized $k$-$\omega$'' (GEKO) two-equation model. The aim of this model is to provide flexibility for the user to fine-tune different coefficients for a RANS computation. GEKO has several parameters that can be fine-tuned to increase predictive accuracy. There are six parameters that can be tuned by the user. In this paper, the focus is on fine-tuning the model coefficients $C_\text{SEP}$ and $C_\text{NW}$ in order to produce more accurate predictions for flow in a converging-diverging channel. The exact equations by which these GEKO coefficients modify the underlying two-equation turbulence closure model are not provided by Ansys, Inc.; however, general guidance for their use is provided in the Ansys User Manual. 
The parameter $C_\text{SEP}$ modifies the characteristics of a separation in the flow. According to the Ansys User Manual, ``increasing $C_\text{SEP}$ reduces eddy viscosity, which leads to more sensitivity to adverse pressure gradients for boundary layers and lower spreading rates for free shear flow” \citep{GEKO_model}. A change in the value of $C_\text{SEP}$ has an influence on all classes of flows. The parameter $C_\text{NW}$ modifies the near-wall characteristics of the flow that affect the nature of the wall boundary layer. The parameter $C_\text{NW}$ generally has little or no effect on free shear flows. An increase in the value of $C_\text{NW}$ leads to a larger wall shear stress. Here, we focus on calibrating global values for the GEKO model coefficients, although local values can also be accommodated through the application of user-defined functions (UDFs). These coefficients have a specified (recommended) range of values which is documented in the Ansys User Manual~\citep{GEKO_model}. In accordance with this documentation, the model coefficients $C_\text{SEP}$ and $C_\text{NW}$ should take values in the range summarized in Table~\ref{tbl:GEKO}.

\begin{table}[]
\centering
\caption{Recommended range of values for the two GEKO coefficients $C_\text{SEP}$ and $C_\text{NW}$.}
\label{tbl:GEKO}
\begin{tabular}{cccc}
\hline
Parameter & Minimum Value & Maximum Value & Default Value \\ \hline
$C_\text{SEP}$ & 0.7           & 2.5           & 1.75    \\
$C_\text{NW}$  & $-2.0$        & 2.0           & 0.50    \\ \hline
\end{tabular}
\end{table}

\subsubsection{Computational setup}
We calculate the steady-state, incompressible, turbulent flow through a converging-diverging channel using the GEKO turbulence closure model in a RANS framework. The DNS Reynolds number for this flow, based on $U_\text{max}$ and $h$, is $Re_{U_\text{max}} = 12,800$. Here, $U_\text{max}$ is the maximum velocity in the fully-developed channel cross-section. The computational domain used for the 2D RANS calculation is shown in Fig.~\ref{fig:domain_channel}. The DNS reference data for this flow was obtained from ~\citet{Marquillie2008, Marquillie2011}. The mesh is adapted from ~\citet{McConkeySciDataPaper2021}, and contains 610,720 grid cells as displayed in Fig.~\ref{fig:mesh_channel}. Mesh independence with the selected mesh density near the converging section was demonstrated in~\citet{McConkeySciDataPaper2021}.

Boundary conditions and fluid properties for the RANS calculation are prescribed to match those of the DNS reference simulation. The inlet plane of the computational domain is placed at a distance of $20h$ upstream of the converging section, where $h$ is the half-channel height. A Neumann pressure boundary condition is imposed at the inlet plane (viz., the gradient of the pressure in the streamwise direction is zero). The uniform streamwise velocity is prescribed at the inlet with a value of 0.845 m~s$^{-1}$. This value was chosen to match the mass flow rate from the original RANS computation undertaken by ~\citet{McConkeySciDataPaper2021}. The inlet plane turbulent kinetic energy value is set to $k = 4.28421 \times 10^{-4}$  m$^2$~s$^{-2}$, and specific dissipation rate at the inlet plane is $\omega = 0.26993 \ \text{s}^{-1}$, in conformance with the computations conducted by ~\citet{McConkeySciDataPaper2021}. A fixed gauge pressure of zero~Pa is imposed at the outlet plane of the computational domain, where all other flow quantities are zero-gradient. The outlet plane is placed at a distance of $20h$ downstream of the bump. Lastly, a no-slip boundary condition was applied on all walls in the computational domain.

The fluid properties of the flow were specified as follows: density $\rho= 1 \ \text{kg} \text{ m}^{-3}$ and kinematic viscosity $\nu = 7.9365 \times 10^{-5} \  \text{m}^2 \text{ s}^{-1}$. The SIMPLE-consistent (SIMPLEC) algorithm was used for the pressure-velocity coupling. Convective terms in the momentum equation were discretized with a second-order upwind scheme. The convective terms in the $k$- and $\omega$-transport equations were discretized with a first-order upwind scheme. The diffusion terms in the momentum and turbulence transport equations were discretized with a second-order central difference scheme.

\subsubsection{Optimization problem}
For the converging-diverging channel case, we use the GEDCP objective function to calibrate the model coefficients based on sparse reference data. Specifically, we calibrate $C_\text{SEP}$ and $C_\text{NW}$ based on a sparse set of measurements of the pressure coefficient along the bottom wall of the channel. The pressure coefficient is determined as follows:
\begin{equation}\label{eq:cp}
C_p = \frac{P-P_0}{\tfrac{1}{2}\rho U_\text{max}^2}\ .
\end{equation}
Here, $P$ is the pressure and $P_0$ is the pressure taken at a point downstream of the converging-diverging channel (approximately at a downstream distance of $12h$ from the start of the converging section).

No integral parameters are included in the objective function. The objective function used here is given by
\begin{align}\label{eq:gedcp_cndv}
    f_\text{GEDCP} (C_\text{SEP}, C_\text{NW}) &= - (E_\text{F})(1+\tfrac{1}{2}p)\big|_{C_\text{SEP}, C_\text{NW}}\ ,\\
    E_{\text{F}}(C_\text{SEP}, C_\text{NW}) &=  \text{MAPE}(C_p) \nonumber\\&=  \frac{1}{N_{C_p}}\sum_{i = 1}^{N_{C_p}} \left| \frac{C_{p,i} - \tilde C_{p,i} (C_\text{SEP}, C_\text{NW})}{C_{p,i}} \right|\ .\\
\end{align}
Ten points along the bottom wall have been selected as sparse wall measurements ($N_{C_p}=10$) used as the reference data in the calibration of the model coefficients. These points are indicated by the blue scatter points in Fig.~\ref{fig:cndv_cp}. The misfit error term in the GEDCP objective function is the MAPE in $C_p$ over the ten reference data points. For the upper and lower bounds of $C_\text{SEP}$ and $C_\text{NW}$, we use the Ansys recommended limits of $0.75 \leq C_\text{SEP} \leq 2.5$ and $-2 \leq C_\text{NW} \leq 2$, as discussed previously in Section~\ref{sec:cndv_setup}.

\subsubsection{Results}
A Python script was created to modify and execute an Ansys Workbench journal file. A fixed computational budget of 30 iterations was used. The first iteration corresponds to the default values of the model coefficients. Subsequently, nine additional Sobol-sampled iterations are executed. The remaining 20 iterations use the Bayesian optimization schema. After 30 iterations, the recommended values for $C_\text{SEP}$ and $C_\text{NW}$ are those that maximize the objective function: $C_\text{SEP}=0.7$, and $C_\text{NW}=1.778$.

\begin{figure}
    \centering
    \includegraphics{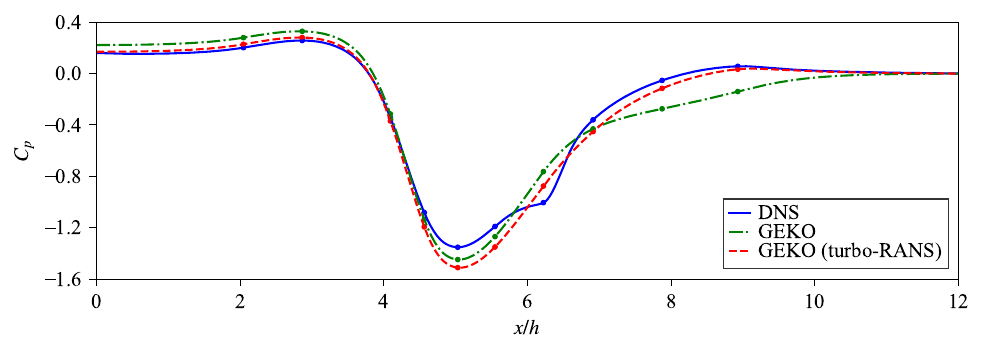}
    \caption{Pressure coefficient along the bottom wall for the converging-diverging channel case. The blue data points exhibit the locations of the measured pressure coefficients that form the reference data used in the calibration of the GEKO turbulence closure model.}
    \label{fig:cndv_cp}
\end{figure}

Fig.~\ref{fig:cndv_cp} compares the DNS data to the predictions obtained using the default and turbo-RANS-optimized GEKO models. After optimization, the streamwise profile of the pressure coefficient is in better conformance with the DNS data. In particular, in the diverging section of the channel, the turbo-RANS-optimized GEKO model better predicts the pressure recovery, in contrast to the default GEKO model which is seen to significantly under-predict the recovery. The optimized model exhibits a slightly greater error in its prediction of the minimum pressure at the top of the converging section (viz., at $x/h\approx 5$), but achieves greater predictive accuracy in every other region of the flow.

\subsection{Discussion}
The preceding three examples demonstrate the flexibility of the GEDCP objective function to work with various reference data types for the calibration of turbulence closure models. A mixture of integral, dense, and sparse reference data was used in these examples. The GEDCP objective function also permits the use of a mixture of reference data types for a given problem. For example, the airfoil case can be modified to include surface pressure coefficient reference data for use in calibration. As demonstrated by the case of the periodic hills, improving predictions of specific quantities of interest often results in improved predictions of other flow quantities (not used as reference data for the calibration). Moreover, when the GEDCP objective function is used as a calibration target in turbo-RANS, a computationally efficient optimization of turbulence model coefficients is possible. Sobol sampling, which occurred until iteration 5 for the airfoil case and iteration 10 for the periodic hills and converging-diverging channel cases, proved to be a highly effective sampling methodology. During the Sobol sampling period, a point close to the optimal value was almost always identified, and therefore the Bayesian optimization algorithm was well-positioned to determine whether to exploit this region or to explore the more sparsely sampled (albeit more uncertain) regions in the parameter space. For more complex flows, there is no guarantee that the Sobol sampling algorithm will happen to provide a point near the optimal value, and exploration within the Bayesian optimization loop may become more important. Section~\ref{sec:hyperparameters} will investigate and provide recommendations for the number of points to be sampled in the initialization phase of Bayesian optimization.

\section{Hyperparameter optimization}\label{sec:hyperparameters}

Complete details of the hyperparameter optimization study are provided in the supplementary information files. In summary, the purpose of the hyperparameter optimization procedure was to determine optimal settings for Bayesian optimization for typical turbo-RANS objective functions. The hyperparameter search spaces are summarized in Table~\ref{tbl:gridsearch}, with each hyperparameter combination being evaluated using 30 different random seeds. Each evaluation consisted of a ``simulated optimization run", which used a set of pre-determined RANS calculations, rather than conducting a new RANS calculation for each coefficient suggestion. A run was considered convergent based on proximity to the optimal solution from the pre-determined set of RANS calculations. In total, 7,110 turbo-RANS optimization loops were simulated. The recommended hyperparameters for turbo-RANS are summarized in Table~\ref{tbl:hyperparameter_recommendation}.

\begin{table}[]
\centering
\caption{Hyperparameter search space.}
\label{tbl:gridsearch}
\begin{tabular}{ccc}
\hline
Hyperparameter   & Values tested                                    & Notes                                                                 \\ \hline
Utility function & UCB, EI, POI                                     & See Section~\ref{sec:utility}for more details.                             \\
$\kappa$ (UCB)   & $0.5 \leq \kappa \leq 6$                         & Linear spaced interval, increment of $\Delta \kappa =0.1$ (56 values). \\
$\xi$ (EI, POI)  & $0 \leq \xi \leq 0.5$                            & Logarithmic spacing, with 56 values.                                   \\
$\nu$            & $\tfrac{1}{2}, \tfrac{3}{2},\tfrac{5}{2},\infty$ & See Section~\ref{sec:gp} for more details.                                             \\
$l$              & $0.01 \leq l \leq 1.0$                           & 14 values total, more concentrated around 0.1, then every 0.1 to 1.0.  \\
$n_s$            & $3\leq n_s \leq 15$                              & 13 values total, every integer.                                        \\ \hline
\end{tabular}
\end{table}

\begin{table}[]\caption{Recommended hyperparameter settings.}\label{tbl:hyperparameter_recommendation}
\centering
\begin{tabular}{ccc}
\hline
Hyperparameter   & Recommended setting                                                    & Notes                                                                                                                                                                                     \\ \hline
Utility function & \begin{tabular}[c]{@{}c@{}}Upper confidence bound\\ (UCB)\end{tabular} & \begin{tabular}[c]{@{}c@{}}Expected improvement (EI) also performs well,\\ but use $\xi \lessapprox 10^{-3}$.\end{tabular}                                                                \\ \hline
$\kappa$         & 2.0                                                                    & \begin{tabular}[c]{@{}c@{}}Minor efficiency gains can be made by lowering $\kappa$,\\ but at the risk of becoming stuck.\end{tabular}                                                     \\ \hline
\rule{0pt}{5ex} $\nu$            & $\dfrac{5}{2}$                                                          & $\nu = \dfrac{3}{2}$ also performs well.                                                                                                                                                  \rule[-4ex]{0pt}{2ex} \\ \hline
$l$              & 0.1                                                                    & \begin{tabular}[c]{@{}c@{}}Since this is just the initial guess for the Gaussian process optimizer,\\ the choice for $l$ is often not significant.\end{tabular}                           \\ \hline
$n_s$            & 10                                                                     & \begin{tabular}[c]{@{}c@{}}For very simple optimization problems (e.g. single variable,\\ under 10 iterations total), $n_s$ can be lowered,\\ at the risk of becoming stuck.\end{tabular} \\ \hline
\end{tabular}
\end{table}

The optimal hyperparameter settings likely change on a case-by-case basis, since the highly customizable GEDCP objective function varies depending on the type of reference data used in the calibration and on the default coefficient preference. Nevertheless, the two relatively different objective functions [Eqs.~(\ref{eq:gedcp_airfoil}) and ~(\ref{eq:gedcp_phll})] have similar optimal hyperparameters. Moreover, the UCB utility function is relatively insensitive to $\kappa$ for these typical objective functions. As the objective function increases in complexity and the number of optimized coefficients increases, the optimal value of $\kappa$ changes slightly but remains close to the recommended default value of $\kappa=2.576$ in the \texttt{bayesian-optimization} library. This value of $\kappa$ represents a reasonable exploration and exploitation tradeoff for a general case, and the optimal turbo-RANS setting of $\kappa=2$ indicates that a slight preference for exploitation is favourable for turbo-RANS. For different flows, settings in the GEDCP function can depend on the number of coefficients in the problem, so the core optimization problem can vary greatly. While we demonstrated here that the recommended setting (UCB, $\kappa=2$) performs well for different problems, it may not perform well for all cases. The Bayesian optimization hyperparameters should be re-tuned if the convergence speed is unsatisfactory for a different objective function.

\section{Conclusion}\label{sec:conclusion}
This work focused on an important and widespread optimization problem in engineering flow simulation: calibrating turbulence model coefficients. The major contributions of this work are to propose a framework for Bayesian optimization of turbulence model coefficients (turbo-RANS), propose a general objective function, demonstrate the predictive accuracy improvements that can be achieved, and recommend hyperparameters for future users of turbo-RANS. We presented the turbo-RANS framework, demonstrated how RANS models can be calibrated using various types of reference data to produce more accurate predictive results, and performed a systematic hyperparameter tuning for the Bayesian optimization process.

Even with perfectly calibrated model coefficients, there is an upper limit to the predictive accuracy that can be achieved with RANS. Errors associated with unavoidable assumptions and approximations (model errors) in RANS can severely degrade the accuracy of these predictions. For example, the assumptions of locality can be violated for many real-world flows~\citep{Pope2000}. ~\citet{Duraisamy2019a} list four main sources of error in RANS: errors arising from ensemble averaging (L1), errors in the functional representation of the Reynolds stress (L2), errors in the model functional form (L3), and errors in the model coefficients (L4). It is possible that for certain classes of flows, the calibration of model coefficients (which reduces the L4 error) can help offset some L2 and L3 errors. However, this effect is difficult to quantify. Nevertheless, even with perfectly calibrated model coefficients, turbo-RANS is only as good as the ``best RANS" solution for a particular problem, which is often limited by unrecoverable L1 and L2 errors. Although turbo-RANS can potentially improve the accuracy of a RANS prediction to some upper bound limit in their predictive accuracy for a given flow, this upper bound limitation in the accuracy will always exist due to implicit modelling assumptions and approximations used in RANS, which may not hold true for many industrially relevant (complex) flows.

Within the wider context of modern machine learning techniques for RANS, we believe that an important future research direction is comparing compare coefficient calibration to highly sophisticated data-driven algebraic Reynolds stress closure models. Although the process of calibrating model coefficients is comparatively simpler than training a non-linear eddy viscosity model, the resulting performance in the present work appears to compete with these machine learning models. Previous results for the periodic hills case by ~\citet{Wu2018,McConkey2022,Kaandorp2020} show similar predictive accuracy to the presently calibrated linear eddy viscosity models. We demonstrated here that simply by tuning model coefficients, there is significant performance to be leveraged from the physics represented in the RANS turbulence closure models. Even further performance can be leveraged from the Bayesian optimization of zonal coefficient values. Currently, Bayesian inference has been applied to infer local turbulence model coefficients, but Bayesian optimization applied to local or zonal model coefficients has not been explored to the authors' knowledge. 

Despite years of optimization methods being applied to turbulence model calibration, and despite the significant accuracy improvements that are possible, these methods are not yet widely used in industrial settings. We believe this may be due to the complicated implementation and expert background required to perform Bayesian calibration. Therefore, by designing an easy-to-use software package, we hope to help spur industrial interest in this technique. All data and code from the present work can be found in the turbo-RANS GitHub repository~\citep{turborans_github}. We also include the 900 periodic hills RANS calculations used for the present hyperparameter tuning investigation, as broader investigations concerned with uncertainty quantification may find this data set useful.

Future work will include model coefficient calibration for more challenging flows. The presently proposed GEDCP objective function can be immediately applied to quantify error in an unsteady quantity of interest, such as a Strouhal number, time-averaged force coefficient, or root-mean-square (RMS) force coefficient. We also aim to explore optimal turbo-RANS hyperparameter tuning for more complex 3D flows and investigate calibrating a RANS turbulence model on a mixture of experimental and numerical reference data. turbo-RANS can also be extended to calibrate coefficients within a Reynolds-stress transport model, or other turbulence closure relations. Ultimately, the algorithm and GEDCP objective function proposed here can be applied to calibrate any coefficient of interest in a RANS model.

Currently, CFD best practices such as a mesh independence study, careful numerical scheme selection, designing meshes appropriate for wall treatment, verification, and validation are missing an important step: coefficient calibration. Previously, this would have been a largely manual activity. However, turbo-RANS automates this activity in a highly computationally efficient manner. We envision that, along with validation, calibration can become a more frequently used practice in RANS. With the more widespread use of closure coefficient calibration, future RANS practitioners can potentially generate a coefficient database, and make an educated selection of coefficients to use in their specific application based on the nature of the physics present in their flow. To some extent, this is the idea behind the GEKO turbulence closure model~\citep{GEKO_model}. However, the GEKO model still requires user tuning of the model coefficients. With sufficient reporting and the use of calibrated RANS models, a recommendation engine (e.g., a large language model) can be embedded into CFD software of the future. This model can be trained to recommend turbulence model coefficients given a description of the flow physics present and quantities of interest. Such a model relies on more reporting of calibrated closure coefficients, which we aim to make easier with turbo-RANS.

\printbibliography
\end{document}


\maketitle

\section{Hyperparameter optimization}\label{sec:hyperparameters}
The objective of hyperparameter optimization was to determine the optimal settings for Bayesian optimization in the turbo-RANS framework. In order to recommend suitable hyperparameters for typical applications involving the calibration of model coefficients of RANS turbulence closure models, two cases were investigated in this context: namely, flow over an airfoil; and, flow over periodic hills. The objective functions for the airfoil and periodic hills cases are given in the main article.

Several optimization runs were completed for these cases with various hyperparameter settings to determine which settings resulted in a solution using the smallest computational effort (i.e., the strategy that arrived at the solution using the smallest number of evaluations of the objective function). Ultimately, the selected hyperparameters allow the optimal RANS model coefficients to be obtained with the lowest computational effort compared to other settings. The following hyperparameters were optimized in this study.
\begin{itemize}
    \item \textbf{Utility function}. Three utility functions are included in the \texttt{bayesian-optimization}~\citep{BayesOptPython} Python library: upper confidence bound (UCB),  expected improvement (EI), and probability of improvement (POI). See the main article for further explanation of these utility functions.
    \item \textbf{Exploration/exploitation tradeoff parameter}. The UCB function is parameterized by a tradeoff parameter $\kappa$, and both the EI and POI functions are parameterized by a tradeoff parameter $\xi$. The default value for the tradeoff parameter in \texttt{bayesian-optimization} for the UCB utility function is $\kappa=2.576$. For both the EI and POI utility functions, the default value for the tradeoff parameter in \texttt{bayesian-optimization} is $\xi=0$. 
    \item \textbf{Gaussian process kernel parameters}. The Matern kernel is used in the underlying Gaussian process regression in turbo-RANS. Important parameters for this kernel are the characteristic length scale $l$ and the parameter $\nu$ (see the main article for more details).
    \item \textbf{Number of initially sampled points}. Before calling the Bayesian optimization, it is important to condition the underlying Gaussian process regression surrogate model of the objective function on some initial points. We use Sobol sampling to randomly generate these points, which provide a uniform sampling of the parameter space~\citep{Sobol1967}.
\end{itemize}
The following procedure was used to select optimal hyperparameters:
\begin{enumerate}
    \item Fix $l$ = 0.1, $\nu = 5/2$, $n_s=5$ for the airfoil case, and $n_s=10$ for the periodic hills case. 
    \item Perform a grid search over both the utility function type and corresponding exploration/exploitation tradeoff parameter.
    \item Select and fix the utility function and exploration/exploitation tradeoff parameter.
    \item Perform a grid search over $l$ and $\nu$.
    \item Select and fix values for $l$ and $\nu$.
    \item Perform a grid search over $n_s$.
\end{enumerate}

\begin{table}[]
\centering
\caption{Hyperparameter search space.}
\label{tbl:gridsearch}
\begin{tabular}{ccc}
\hline
Hyperparameter   & Values tested                                    & Notes                                                                 \\ \hline
Utility function & UCB, EI, POI                                     & See  the main article for discussion.                       \\
$\kappa$ (UCB)   & $0.5 \leq \kappa \leq 6$                         & Linear spaced interval, increment of $\Delta \kappa =0.1$ (56 values) \\
$\xi$ (EI, POI)  & $0 \leq \xi \leq 0.5$                            & Logarithmic spacing, with 56 values                                   \\
$\nu$            & $\tfrac{1}{2}, \tfrac{3}{2},\tfrac{5}{2},\infty$ & See  the main article for discussion.                                            \\
$l$              & $0.01 \leq l \leq 1.0$                           & 14 values total, more concentrated around 0.1, then every 0.1 to 1.0  \\
$n_s$            & $3\leq n_s \leq 15$                              & 13 values total, every integer                                        \\ \hline
\end{tabular}
\end{table}
The grid search space for each parameter is summarized in Table~\ref{tbl:gridsearch}. In total, 237 hyperparameter combinations were tested. Given the random nature of the Gaussian process regression that underpins Bayesian optimization, the core turbo-RANS algorithm is not deterministic. However, the algorithm can be made deterministic in practice by setting the random seed in the code. To better select optimal hyperparameters, 30 random turbo-RANS optimization runs were completed for each hyperparameter setting, for a total of 7,110 optimization runs. Following this, the mean and standard deviation of the number of runs to reach convergence were determined.  The mean number of runs, $\overline{n}$, is determined in accordance to
\begin{equation}\label{eq:mean}
    \overline{n} = \frac{1}{N}\sum_{i=1}^N n_i\ ,
\end{equation}
where $N$ is the total number of turbo-RANS runs, $i$ indexes each turbo-RANS run, and $n_i$ is the number of iterations to reach convergence in the $i$-th run. The standard deviation, $s$, of the number of runs required to achieve convergence is given by
\begin{equation}\label{eq:stdev}
     s = \left(\frac{1}{N}\sum_{i=1}^N \bigl(n_i - \overline{n}\bigr)^2\right)^{1/2}\ .
\end{equation}

The purpose of applying turbo-RANS is to tune turbulence model coefficients based on a case-specific objective function. To accelerate the hyperparameter tuning, the objective functions for the airfoil case and periodic hills case were evaluated \textit{a priori} over a series of $a_1$ values for the airfoil case, and a grid of values of $(a_1, \beta^*)$ for the periodic hills case. These \textit{a priori} simulations permitted faster hyperparameter optimization since the value of $f_\text{GEDCP}$ given a model coefficient suggestion can be determined by interpolation. This \textit{a priori} evaluation of $f_\text{GEDCP}$ on a grid of values of the model coefficients would not be used in practice with turbo-RANS. In a typical application, a new RANS prediction would need to be computed with each model coefficient suggestion. In other words, since the hyperparameter optimization procedure required thousands of evaluations of $f_\text{GEDCP}$, it was more computationally efficient to ``simulate a simulation" by interpolating between a set of previously computed RANS calculations. These \textit{a priori} RANS calculations also allow visualization of the objective function. The objective function for the airfoil case is shown in the main article for a set of 50 RANS calculations for values of $a_1$ that span the parameter space. Similarly, the objective function for the periodic hills case is shown in the main article for a set of 900 RANS calculations for combinations of values of the two model coefficients $(a_1,\beta^*)$ that span the parameter space.

Convergence criteria in turbo-RANS are determined by the user. For the hyperparameter optimization, the optimal coefficients and values of $f_\text{GEDCP}$ were known because a large number of \textit{a priori} RANS calculations were completed on a grid. For the airfoil case, the optimization procedure was considered convergent when the suggested value of the $a_1$ coefficient was within $0.5\%$ of the optimal value of this coefficient. For the case of the periodic hills, the optimization procedure was considered convergent when the suggested values for the coefficients $(a_1,\beta^*)$ resulted in an objective function $f_\text{GEDCP}$ value that was within $5\%$ of the optimum objective function value. The choice of convergence criteria for hyperparameter optimization is a special case since it relies on \textit{a priori} knowledge of the objective function optimum. In practice, turbo-RANS convergence criteria can be set based on the change in the best objective function or best coefficients over a given window of iterations. Given the expensive nature of RANS calculations, we envision that many turbo-RANS optimization procedures will be run with a prescribed computational budget instead of a numerical convergence criteria (e.g., a 40-simulation optimization run can be prescribed {\it a priori}). The Bayesian optimization algorithm produces a computationally efficient result for either of these criteria.

\subsection{Utility function}
\begin{figure}
     \centering
     \begin{subfigure}[c]{0.9\textwidth}
        \centering
        \includegraphics{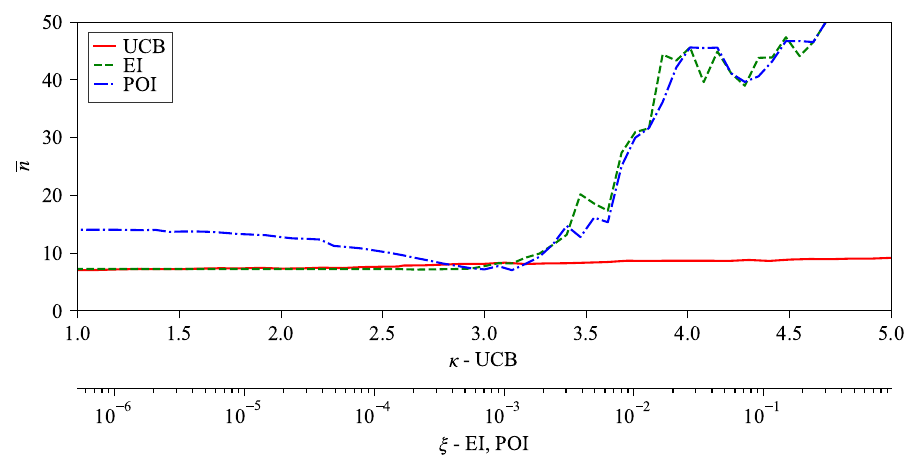}
        \caption{Comparison of the mean value $\overline{n}$ of the number of runs required to achieve convergence as a function of the exploration/exploitation tradeoff parameter for the UCB, EI, and POI utility functions.}
     \end{subfigure}
     \begin{subfigure}[b]{0.3\textwidth}
         \centering
         \includegraphics{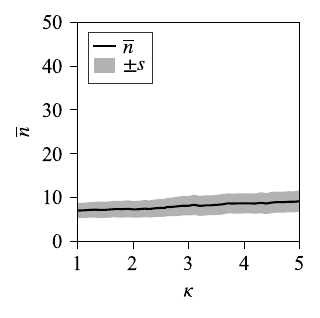}
         \caption{Mean $\overline{n}$ and standard deviation $s$ of the number of runs required to achieve convergence as a function of the exploration/exploitation tradeoff parameter $\kappa$ for the UCB utility function.}
     \end{subfigure}
     \hfill
     \begin{subfigure}[b]{0.3\textwidth}
         \centering
         \includegraphics{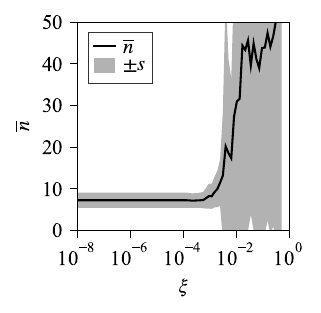}
         \caption{Mean $\overline{n}$ and standard deviation $s$ of the number of runs required to achieve convergence as a function of the exploration/exploitation tradeoff parameter $\xi$ for the EI utility function.}
     \end{subfigure}
     \hfill
     \begin{subfigure}[b]{0.3\textwidth}
         \centering
         \includegraphics{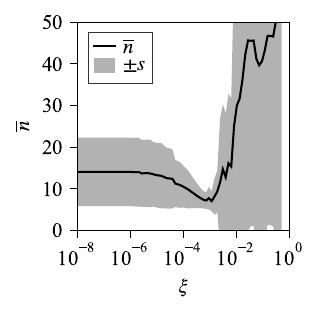}
         \caption{Mean $\overline{n}$ and standard deviation $s$ of the number of runs required to achieve convergence as a function of the exploration/exploitation tradeoff parameter $\xi$ for the POI utility function.}
     \end{subfigure}
        \caption{Utility function hyperparameter search results for the airfoil case.}
        \label{fig:utility_airfoil}
\end{figure}

Fig.~\ref{fig:utility_airfoil} shows the results of the utility function search for the airfoil case. In terms of purely minimizing the number of evaluations of the objective function, the UCB, EI, and POI utility functions all give roughly the same minimum mean number of evaluations of $\overline{n}=7$. For the UCB and EI utility functions, the performance is stable over a wide range of values of $\kappa$ and $\xi$. However, past a critical value of $\xi$, the EI utility function becomes too explorative, and the number of objective function evaluations for convergence increases sharply. The UCB utility function achieves stable performance for all values of $\kappa$, with a minor trend of an increasing $\overline{n}$ with an increasing value of $\kappa$. The POI utility function has an optimum $\xi$ value. Below this optimum value, the performance of the POI utility function is worse than that of the UCB or EI utility functions. Above this optimum value, the POI function becomes highly explorative, performing similarly to the EI utility function. Furthermore, the standard deviation of the 30 random runs is consistent for the UCB utility function, indicating that all the runs achieve convergence in a similar number of iterations. When the EI and POI utility functions become overly explorative, the standard deviation is seen to increase sharply. 

\begin{figure}
     \centering
     \begin{subfigure}[c]{0.9\textwidth}
        \centering
        \includegraphics{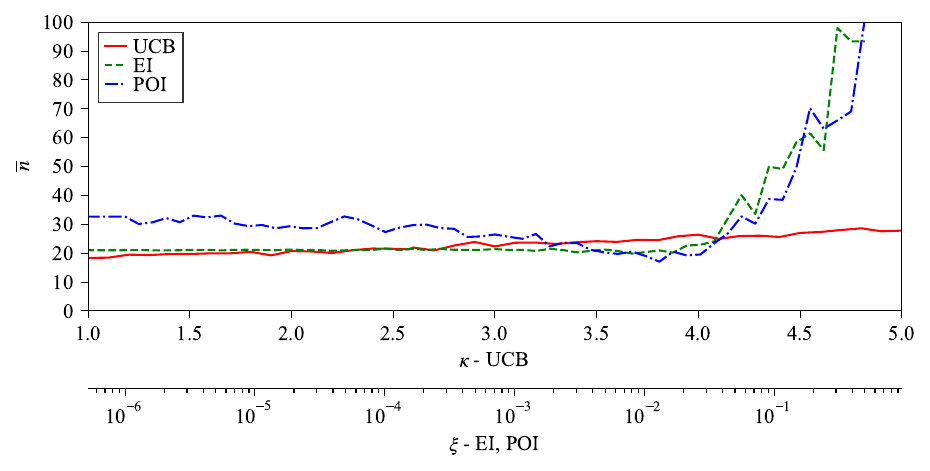}
        \caption{Comparison of the mean value $\overline{n}$ of the number of runs required to achieve convergence as a function of the exploration/exploitation tradeoff parameter for the UCB, EI, and POI utility functions.}
     \end{subfigure}
     \begin{subfigure}[b]{0.3\textwidth}
         \centering
         \includegraphics{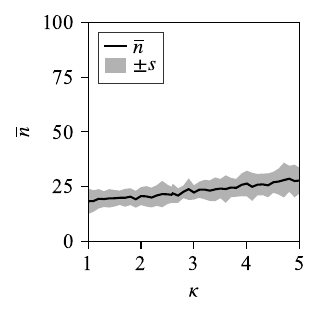}
         \caption{Mean $\overline{n}$ and standard deviation $s$ of the number of runs required to achieve convergence as a function of the exploration/exploitation tradeoff parameter $\kappa$ for the UCB utility function.}
     \end{subfigure}
     \hfill
     \begin{subfigure}[b]{0.3\textwidth}
         \centering
         \includegraphics{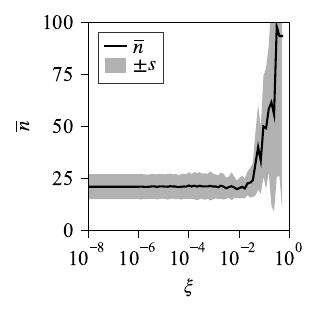}
         \caption{Mean $\overline{n}$ and standard deviation $s$ of the number of runs required to achieve convergence as a function of the exploration/exploitation tradeoff parameter $\xi$ for the EI utility function.}
     \end{subfigure}
     \hfill
     \begin{subfigure}[b]{0.3\textwidth}
         \centering
         \includegraphics{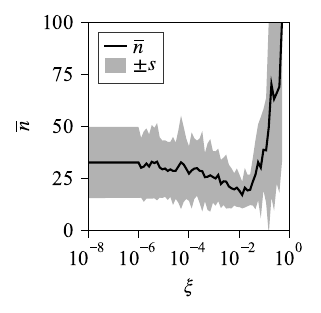}
         \caption{Mean $\overline{n}$ and standard deviation $s$ of the number of runs required to achieve convergence as a function of the exploration/exploitation tradeoff parameter $\xi$ for the POI utility function.}
     \end{subfigure}
        \caption{Utility function hyperparameter search results for the periodic hills case.}
        \label{fig:utility_phll}
\end{figure}

Fig.~\ref{fig:utility_phll} shows the hyperparameter optimization results for the periodic hills case. The general trends are similar to the airfoil case: the UCB utility function performs consistently, and the EI and POI functions have a critical $\xi$ value beyond which the Bayesian optimization process becomes overly explorative. Similar trends to the airfoil case are also seen for the standard deviation. Given the two-coefficient optimization, a larger coefficient search space, and a more complex objective function, it is not surprising that the periodic hills case requires more iterations to find the optimal point than the airfoil case, with the minimum mean number of objective function evaluations being 20. 

Comparing Figs.~\ref{fig:utility_airfoil} and~\ref{fig:utility_phll}, we see that, when optimized, all three utility functions perform equally well. However, a major disadvantage of the POI utility function is the need to determine the optimum $\xi$, which is problem-specific: it is $\xi \approx 10^{-3}$ for the airfoil case and $\xi\approx 10^{-2}$ for the periodic hills case. Given their similar performance and low standard deviation, the EI and UCB utility functions are both acceptable choices for use in the acquistion of a new point in the Bayesian optimization process. However, the EI utility function will rapidly become overly explorative past a problem-dependent threshold. Therefore, for a general turbo-RANS use case, we recommend the application of the UCB utility function.

The optimal exploration/exploitation tradeoff parameter for the GEDCP objective function generally favours exploitation. However, when the $\kappa$ or $\xi$ parameters were too small (overly promoting exploitation), we observed that the Bayesian optimization process became ``stuck'' on a perceived local maximum. For the UCB utility function, the Bayesian optimization process can occasionally become stuck for values of $\kappa \lessapprox 1$. Indeed, these runs would not converge after hundreds of iterations, and therefore they were removed from the mean and standard deviation calculations. It is noted that these ``stuck" runs are indicative of overly exploitative behavior. Therefore, we recommend a value of $\kappa = 2$, so that the optimization is slightly more (but not overly) exploitative than the default value for $\kappa$ recommended in \texttt{bayesian-optimization}.

A strategy that can be used to ensure that the Bayesian optimization process does not become ``stuck'' at a particular query point selected by the utility function is to simply add some ``noise'' (jitter) to a selected query point. This will guarantee that the selected query point in the current iteration of the Bayesian optimization is different from a previously selected query point. We have not investigated this option in the current study, but it is a possible extension that will be considered in some future work on turbo-RANS.

\subsection{Matern kernel parameters}
\begin{figure}
     \centering
     \begin{subfigure}[b]{0.3\textwidth}
        \centering
        \includegraphics{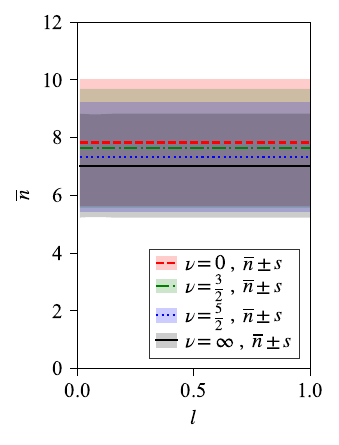}
        \caption{Airfoil case.}
     \end{subfigure}
     \hfill
     \begin{subfigure}[b]{0.6\textwidth}
         \centering
         \includegraphics{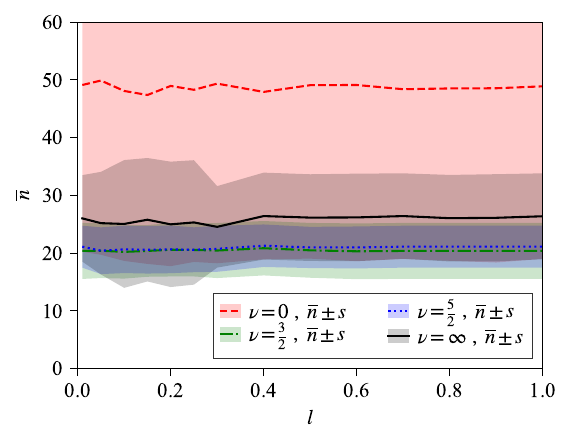}
         \caption{Periodic hills case.}
     \end{subfigure}
        \caption{Mean $\overline{n}$ and standard deviation $s$ of the number of runs required to achieve convergence for the Matern kernel for different values of the hyperparameter $\nu$.}
        \label{fig:kernel_hyperparams}
\end{figure}

After selecting the UCB utility function with $\kappa=2.0$, the Matern kernel parameters were optimized. A grid search was completed for values of $\nu$ and $l$ in the following ranges: $\nu \in [0,\infty)$ and $l \in (0,1]$. Fig.~\ref{fig:kernel_hyperparams} shows the results of this grid search. Neither the airfoil nor the periodic hills cases show sensitivity to the length scale value. Performance in the airfoil case is not very sensitive to the $\nu$ value, with all values of $\nu$ achieving convergence of between 7 and 8 objective function evaluations on average. For the periodic hills case, it becomes clear that $\nu=0$ does not perform well, whereas $\nu={3}/{2}$ and ${5}/{2}$ achieve similar performance. We recommend setting $\nu={5}/{2}$, which is the default value recommended for the \texttt{scikit-learn} Matern kernel~\citep{scikit-learn}. This kernel returns a twice-differentiable function. For the length scale, we recommend a value of $l=0.1$. This length scale value specifies the initial guess for conditioning the Gaussian process, and the final length scale inside the kernel is optimized internally in \texttt{scikit-learn}. Therefore, only a reasonable initial guess for this hyperparameter needs to be specified. For optimizing the Gaussian process parameters, using a sufficient number of sample points initially before the Bayesian optimization was much more important than the choice of the initial value for the length scale.

\subsection{Number of initial samples}
After selecting $\nu={5}/{2}$ and $l = 0.1$, the last hyperparameter to be optimized was the number of initial samples, $n_s$. As discussed in the main article, a certain number of points within the parameter space are obtained using Sobol sampling before calling the Bayesian optimization algorithm. The purpose of these points is to provide sufficient information to properly perform Gaussian process regression on the objective function. When $n_s$ is too small, errant behavior of the Gaussian process can occur. This errant behavior results in subsequent slow optimization, with the Bayesian optimization process often getting immediately stuck. When $n_s$ is too large, then the overall algorithm becomes computationally inefficient by overly sampling the parameter space prior to entering the efficient Bayesian optimization loop.

\begin{figure}
     \centering
     \begin{subfigure}[b]{0.48\textwidth}
        \centering
        \includegraphics{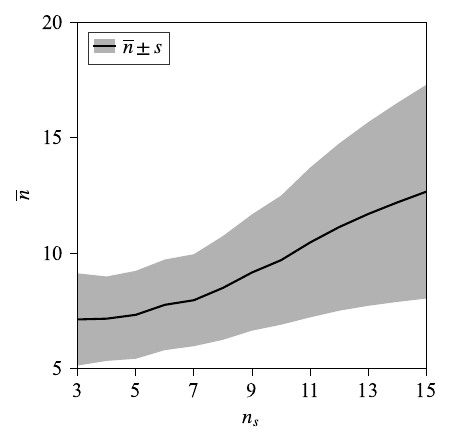}
        \caption{Airfoil case.}
     \end{subfigure}
     \hfill
     \begin{subfigure}[b]{0.48\textwidth}
         \centering
         \includegraphics{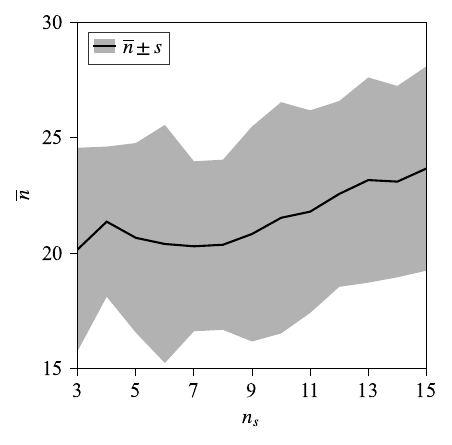}
         \caption{Periodic hills case.}
     \end{subfigure}
        \caption{Mean $\overline{n}$ and standard deviation $s$ of the number of runs required to achieve convergence for various numbers $n_s$ of initial sample points.}
        \label{fig:n_sample}
\end{figure}

Fig.~\ref{fig:n_sample} shows the results of optimizing $n_s$ for the airfoil and periodic hills cases. For the airfoil case, it is optimal to reduce the number of sample points as much as possible. Below $n_s=3$, the previously mentioned erratic behavior can occur, as the Gaussian process regression of the objective function is fit on too few points. For the periodic hills case, there is a local minimum at $n_s\approx8$. These results show that the number of initial samples should be increased as the number of optimized parameters increases. Given our experience concerning the errant behavior of the Gaussian process regression with too few initial samples, a conservative default has been set in the turbo-RANS algorithm of $n_s=10$. If the end-user wishes to increase performance further, we recommend first reducing the value of $n_s$, but generally keeping $n_s>5$ for maintaining the stability of the Gaussian process regression.

\printbibliography